\documentclass[prd, 10pt,nofootinbib, superscriptaddress, preprintnumbers, floatfix]{revtex4}
\usepackage{color}
\usepackage{dsfont}
\usepackage{graphicx}
\usepackage{amssymb, amsfonts, amsmath}
\usepackage{hyperref}
\usepackage{subfigure}
\usepackage[utf8]{inputenc}

\graphicspath{{./}}
\renewcommand{\l}{\left}
\renewcommand{\r}{\right}

\newcommand{\g}[1]{\gamma_{#1}} 

\newcommand{\bra}[1]{\left< #1 \right|} 
\newcommand{\ket}[1]{\left| #1 \right>} 
\newcommand{\gev}{\,\mathrm{GeV}}
\newcommand{\mev}{\,\mathrm{MeV}}
\newcommand{\fm}{\,\mathrm{fm}}
\newcommand{\order}[1]{\mathcal{O}\l({#1}\r)}

\newcommand{\syserr}[1]{(#1)_\mathrm{sys}}
\newcommand{\staterr}[1]{(#1)_\mathrm{stat}}
\newcommand{\tsep}{t_\mathrm{sep}}

\newcommand{\Ctwopt}[2]{C_\mathrm{2pt}(\vec{#2}, #1)}

\newcommand{\qmax}{q_\mathrm{max}}

\newcommand{\qvec}{\vec{q}}

\newcommand{\evec}{\vec{e}}
\newcommand{\khat}{\hat{k}}
\newcommand{\Pibar}{\bar{\Pi}}
\newcommand{\asin}{\mathrm{asin}}
\newcommand{\brackets}[1]{\langle #1 \rangle}

\begin{document}
\title{A model-independent determination of the nucleon charge radius from lattice QCD}

\newcommand\bn{HISKP and BCTP, Rheinische Friedrich-Wilhelms Universit\"at Bonn, 53115 Bonn, Germany}
\newcommand\cyi{Cyprus Computation-based Science and Technology Research Center, 20 C. Kavafi Street, 2121 Nicosia, Cyprus}
\newcommand\mz{PRISMA Cluster of Excellence and Institut f\"ur Kernphysik, Johann-Joachim-Becher-Weg~45, University of Mainz, 55099 Mainz, Germany}
\newcommand\ucy{Department of Physics, University of Cyprus, PO Box 20537, 1678 Nicosia, Cyprus}

\author{Constantia~Alexandrou}\affiliation{\ucy}\affiliation{\cyi}
\author{Kyriakos Hadjiyiannakou}\affiliation{\ucy}\affiliation{\cyi}
\author{Giannis Koutsou}\affiliation{\cyi}
\author{Konstantin Ottnad}\email{kottnad@uni-mainz.de}\affiliation{\mz}
\author{Marcus Petschlies}\affiliation{\bn}

\begin{abstract}
 \begin{center}
  \includegraphics[draft=false,width=.20\linewidth]{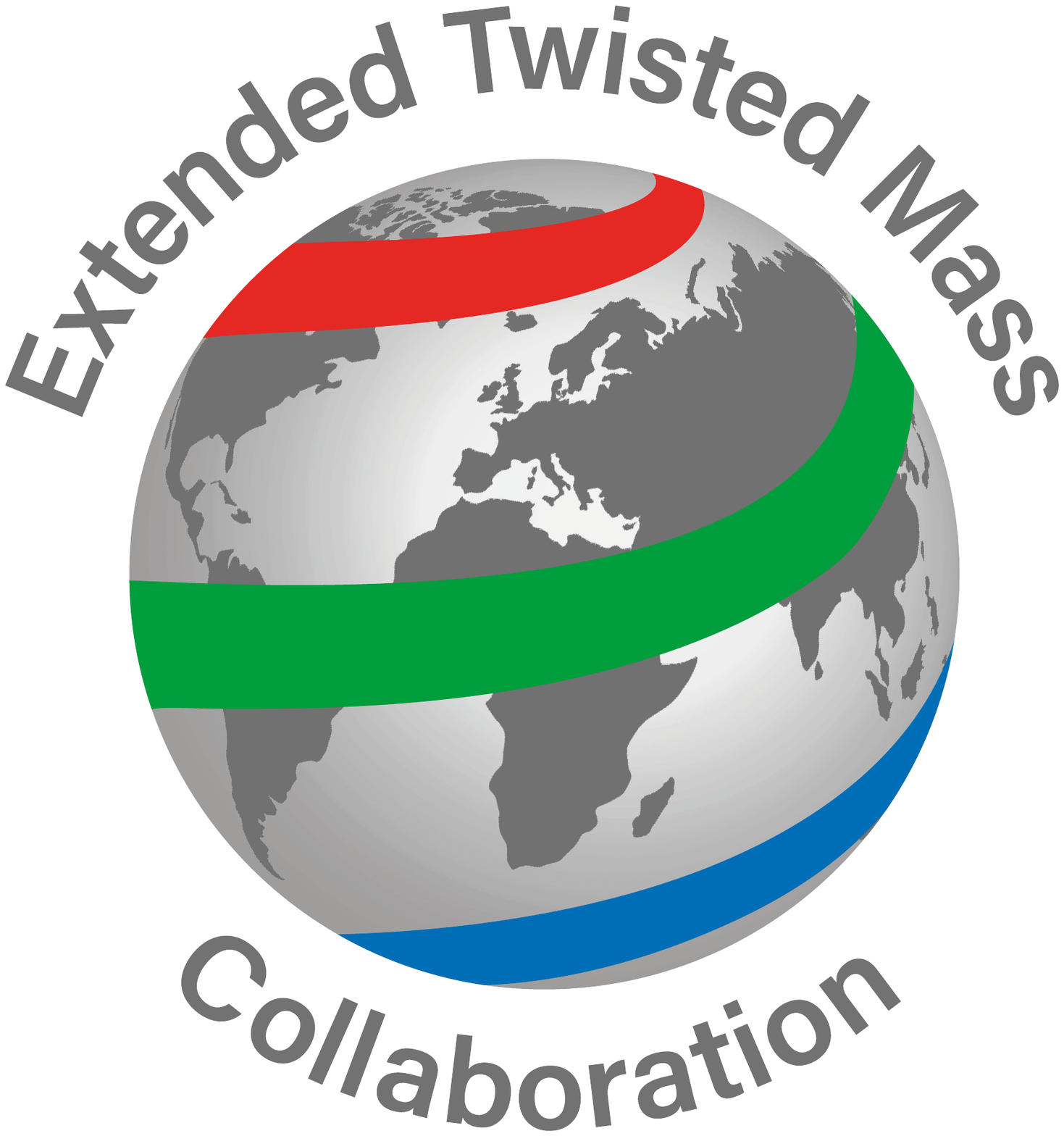}
 \end{center}
  \noindent Lattice QCD calculations of nucleon form factors are restricted to discrete values of the Euclidean four-momentum transfer. Therefore, the extraction of radii typically relies on parametrizing and fitting the lattice QCD data to obtain its slope close to zero momentum transfer. We investigate a new method, which allows to compute the nucleon radius directly from existing lattice QCD data, without assuming a functional form for the momentum dependence of the underlying form factor. The method is illustrated for the case of the isovector mean-square charge radius of the nucleon $\langle r^2_\mathrm{isov} \rangle$ and the quark-connected contributions to $\langle r^2_p\rangle$ and $\langle r^2_n \rangle$ for the proton and neutron, respectively. Computations are performed using a single gauge ensemble with $N_f=2+1+1$ maximally twisted mass clover-improved fermions at physical quark mass and a lattice spacing of $a=0.08\fm$.
\end{abstract}

\maketitle
\clearpage

\section{Introduction} \label{sec:introduction}
The radius of the proton is a fundamental quantity for atomic, nuclear, and particle physics. In atomic physics, it enters in the determination of the Rydberg constant, the most precisely known constant in nature, as well as in precision tests of quantum electrodynamics. In nuclear physics, it characterizes the size of the most abundant hadron in nature and in particle physics it is an input for beyond the standard model physics, testing lepton universality and the possible existence of new particles. Electron-proton scattering was traditionally used to determine the proton radius that is extracted from the slope of the electric Sachs form factor $G_E(q^2)$ extrapolated to zero momentum transfer. The proton radius value extracted from precision measurements of elastic electron scattering cross sections at the Mainz Microtron found a proton radius of $r_p=0.879\staterr{5}\syserr{4}\fm$~\cite{Bernauer:2010wm}, where the systematic errors are added quadratically. Combining electron scattering data with more accurate data from the hyperfine structure of electronic hydrogen, the value recommended by CODATA was $r_p=0.8775(51)$~fm~\cite{Mohr:2012tt}, although a reanalysis of the Mainz electron-proton scattering data found smaller values $r_p=0.84(1)$~\cite{Lorenz:2012tm} and $r_p=0.840^{+15}_{-12}$~\cite{Lorenz:2014yda} compatible with older determinations; cf.~\cite{Mergell:1995bf}. In a pioneering experiment in 2010, the radius of the proton was extracted from the Lamb shift measured for muonic hydrogen~\cite{Pohl:2010zza}. The muonic hydrogen determination is much more accurate and the value extracted $r_p=0.844087(39)$~fm~\cite{Antognini:1900ns} differs by 7 standard deviations from the value recommended by CODATA. This gave rise to the proton radius puzzle. A follow-up, very accurate spectroscopic measurement using electronic hydrogen found a value that was inconsistent with previous determinations and consistent with the size extracted from the muonic hydrogen experiments, namely, $r_p = 0.8335(95)$fm~\cite{Beyer:2017gug}. However, another spectroscopic measurement found a value of $r_p=0.877(13)$~fm\cite{Fleurbaey:2018fih} in agreement with the CODATA value. This discrepancy between the two most recent spectroscopic measurements conducted on electronic hydrogen remains unresolved and is under investigation. However, recently, the proton charge radius experiment at Jefferson Laboratory (PRad), overcoming several limitations of previous electron-proton scattering experiments, enabled measurements at very small forward-scattering angles. Using their results, they extracted a proton radius of $r_p = 0.831\staterr{7}\syserr{12}\fm$~\cite{Xiong:2019umf} in agreement with the muonic hydrogen extraction, toward a possible resolution of the puzzle. We also refer to Ref.~\cite{Hammer:2019uab} for a summary of the current state of this debate.

A theoretical determination of the proton radius starting from the fundamental theory of the strong interaction requires a nonperturbative framework. In this work, we used the lattice QCD formulation to determine the radius of the proton. The standard procedure in lattice QCD is to compute the electric form factor for finite values of the momentum transfer and then perform a fit to determine the slope at zero momentum transfer. However, on a finite lattice, the smallest nonzero momentum is $2\pi/L$ where $L$ is the spatial size of the lattice. Therefore, to reach very small momentum transfers one needs very large lattices. In this work, we use a direct method to extract the proton radius that does not depend on fitting the form factor. We illustrate the validity of our method by analyzing one ensemble of gauge configuration generated at the physical pion mass.

The structure of this paper is as follows: In Sec.~\ref{sec:setup}, we describe the general lattice setup, in Sec.~\ref{sec:method}, we explain our new direct approach to extract the radius, in Sec.~\ref{sec:results}, we discuss our results and in Sec.~\ref{sec:summary}, we compare to other lattice QCD determinations and give our conclusions.

\section{Lattice setup} \label{sec:setup}
Our lattice calculations are performed on a single ensemble with $N_f=2+1+1$ flavors of twisted mass \cite{Frezzotti:2000nk} Wilson clover-improved fermions at maximal twist \cite{Alexandrou:2018egz,Abdel-Rehim:2015pwa}. The quark masses on this ensemble are tuned to their physical masses; however, the simulations do not account for any isospin splitting for the light quarks. Furthermore, the ensemble has been tuned to maximal twist to implement automatic $\mathcal{O}(a)$ improvement \cite{Frezzotti:2003ni,Frezzotti:2003xj}. The gauge ensemble is simulated with a spatial volume of $(L/a)^3=64^3$ and a temporal extent of $T/a=128$. The lattice spacing and pion mass have been determined in Ref.~\cite{Alexandrou:2018egz} with values of $a=0.0809\staterr{2}\syserr{4} \fm$ and $M_\pi=138.0\staterr{4}\syserr{6}\mev$, respectively. The value of $M_\pi$ is very close to its physical value in the isospin limit~\cite{Aoki:2016frl}. For further details on the simulation we refer to Ref.~\cite{Alexandrou:2018egz}. \par

Measurements are performed on 750 gauge configurations, measuring on only every fourth configuration which corresponds to a separation of four hybrid Monte Carlo trajectories between consequent measurements. All statistical errors in our analysis are computed from a binned jackknife procedure taking into account correlations and possible residual affects of autocorrelations.\par

\subsection{Electric form factor of the nucleon}
The aim of this study is an extraction of the nucleon charge radius; hence we consider the electromagnetic matrix element of the nucleon
\begin{equation}
 \bra{N(p_f,s_f)} j_\mu \ket{N(p_i,s_i)} = \bar{u}(p_f, s_f) \left[\g{\mu} F_1(Q^2) + \frac{\sigma_{\mu\nu} Q_\nu}{2m_N} F_2(Q^2) \right]  u(p_i,s_i) \,,
 \label{eq:matrix_element}
\end{equation}
where on the left-hand side $N(p_i,s_i)$ ($N(p_f,s_f)$) label nucleon states with initial (final) state momentum $p_i$ ($p_f$) and spin $s_i$ ($s_f$), and $j_\mu$ is a vector current insertion which will be discussed below. On the right-hand side, $u(p_i,s_i)$, $\bar{u}(p_f, s_f)$ denote Dirac spinors, $m_N$ the nucleon mass, and we have introduced the Dirac and Pauli form factors $F_1(Q^2)$ and $F_2(Q^2)$, which depend on the Euclidean four-momentum transfer $Q^2=-q^2$ with $\vec{q}=\vec{p}_f-\vec{p}_i$. We will work in Euclidean spacetime throughout this study and in the following discussion we will always replace $F_1(Q^2)$ and $F_2(Q^2)$ by the electromagnetic Sachs form factors $G_E(Q^2)$ and $G_M(Q^2)$ which are more convenient for our purposes and related to $F_1(Q^2)$ and $F_2(Q^2)$ via
\begin{align}
 G_E(Q^2) &= F_1(Q^2) - \frac{Q^2}{4m_N^2} F_2(Q^2) \,, \label{eq:G_E} \\
 G_M(Q^2) &= F_1(Q^2) + F_2(Q^2) \,. \label{eq:G_M}
\end{align}
The electromagnetic vector current $j^{em}_\mu(x)$ can be expressed in terms of vector currents $j_\mu^f(x)$ for the individual quark flavors $f$ through
\begin{equation}
 j_\mu^\mathrm{em}(x) = \sum_f e_f j_\mu^f(x) \,,
 \label{eq:elmag_current}
\end{equation}
where $e_f$ denotes the electrical charge for a quark of flavor $f$. On the lattice, we employ the symmetrized conserved vector current which for a quark field $\chi_f(x)$ in the twisted mass basis is given by
\begin{align}
 j_\mu^f(x) = \frac{1}{4} &\l[  \bar{\chi}_f(x+a\hat{\mu}) U^\dag_\mu(x) (1+\g{\mu}) \chi_f(x) - \bar{\chi}_f(x) U_\mu(x) (1-\g{\mu}) \chi_f(x+a\hat{\mu}) \r. \notag \\
                          &\l. +\bar{\chi}_f(x) U^\dag_\mu(x-a\hat{\mu}) (1+\g{\mu}) \chi_f(x-a\hat{\mu}) - \bar{\chi}_f(x-a\hat{\mu}) U_\mu(x-a\hat{\mu}) (1-\g{\mu}) \chi_f(x) \r] \,,
 \label{eq:CVC}
\end{align}
where $U_\mu(x)$ denotes a gauge link and $\hat{\mu}$ a unit vector in the $\mu$ direction. We restrict our calculation to light valence quarks and the relation of the light quark doublet $\chi(x)=(\chi_u(x), \chi_d(x))^T$ to the physical up and down quark at maximal twist is given by the chiral rotation 
\begin{align}
 \chi(x)       &= \frac{1}{\sqrt{2}} \l( 1 + i \g{5} \tau^3\r) \psi(x)  \,, \\
 \bar{\chi}(x) &= \bar{\psi}(x) \frac{1}{\sqrt{2}} \l( 1 + i \g{5} \tau^3\r) \,.
 \label{eq:tm_rotation}
\end{align}

In the $SU(2)$ isospin symmetry limit, the matrix elements of the electromagnetic current satisfy
\begin{equation}
 \bra{p} j_\mu^{em} \ket{p} - \bra{n} j_\mu^{em} \ket{n} = \bra{p} j^u_\mu - j^d_\mu \ket{p} \equiv \bra{p} j^{u-d}_\mu \ket{p}\,,
 \label{eq:pn_jud}
\end{equation}
where the isovector current $j^{u-d}_\mu(x)$ has been introduced and $\ket{p}$, $\ket{n}$ denote proton and neutron states, respectively. Unlike the electromagnetic combination, the isovector current does not give rise to quark-disconnected diagrams in our lattice simulations. However, in this work, we neglect any quark-disconnected contributions as our main goal is to present and benchmark a new method to extract nucleon radii. We remark that these contributions have been shown to be small compared to the quark-connected contribution in Refs.~\cite{Alexandrou:2018sjm,Alexandrou:2019olr}. \par

In the twisted mass lattice regularization, the flavor symmetry is broken from $SU(2)$ to the subgroup $U(1)_3$ and the isospin symmetry-based relations, Eq.~(\ref{eq:pn_jud}), are valid at nonzero lattice spacing up to lattice artifacts. In particular, the quark-disconnected contribution from the insertion of the isovector current $j^{u-d}$ will be discarded from our calculation, since it vanishes in the continuum limit.\par

The point-split vector current in Eq. (\ref{eq:CVC}) is the Noether current associated with the residual flavor symmetry group $U(1)\times U(1)_3$
and thus remains conserved. Therefore, no additional multiplicative renormalization of the matrix element is required. \par

\subsection{Two- and three-point functions}
The lattice QCD evaluation of the nucleon matrix element in Eq.~(\ref{eq:matrix_element}) requires the computation of two- and three-point functions
\begin{align}
 \Ctwopt{t_f-t_i}{p} &= \Gamma_0^{\beta\alpha} \sum_{\vec{x}_f} e^{-i \vec{p} \cdot (\vec{x}_f-\vec{x}_i)} \langle J_{N,\alpha}(\vec{x}_f,t_f) \bar{J}_{N,\beta}(\vec{x}_i,t_i)\rangle \,, \label{eq:2pt_xspace} \\
 C_\mu(\Gamma_\nu, \vec{q}, t_{op}-t_i, t_f-t_i) &= \Gamma_\nu^{\beta\alpha} \sum_{\vec{x}_f, \vec{x}_{op}}  e^{-i \vec{p}_f \cdot (\vec{x}_{f}-\vec{x}_{op})} e^{i \vec{p}_i \cdot (\vec{x}_{op}-\vec{x}_{i})} \langle J_{N,\alpha}(\vec{x}_f, t_f) j^{u-d}_\mu(\vec{x}_{op}, t_{op}) \bar{J}_{N,\beta}(\vec{x}_i, t_i)\rangle \,, \label{eq:3pt_xspace}
\end{align}
where $t_i$, $t_{op}$, and $t_f$ label initial (source), operator insertion, and final (sink) time slices, respectively. The spin projectors $\Gamma_\nu$ are given by $\Gamma_0=\frac{1}{2}(1+\g{0})$ for $\nu=0$ and $\Gamma_k=\Gamma_0 i \g{5}\g{k}$ for $\nu=k=1,2,3$. By time translation invariance, only time differences are of physical relevance and it is convenient to introduce the source-sink time separation $\tsep=t_f-t_i$ and the shorthand $t=t_{op}- t_i$. Our kinematic setup is chosen such that the final state is produced at rest, i.e., $\vec{p}_f=0$, $\vec{p}_i=-\vec{q}$. Finally, the proton interpolating field is given in the physical basis by
\begin{equation}
 J_{N,\alpha}(x) = \epsilon_{abc} \l(u_a^T(x) C \g{5} d_b(x)\r) u_{c,\alpha}(x) \,,
 \label{eq:interpolating_operator}
\end{equation}
where $C$ denotes the charge conjugation matrix and is transformed into the twisted basis in terms of $\chi_{u/d}$ by using the chiral rotation in Eq. (\ref{eq:tm_rotation}). Since nucleon structure calculations are known to be hampered by a severe signal-to-noise problem, it is crucial to increase the overlap of the interpolating operator with the desired nucleon ground state, effectively suppressing excited states and allowing to extract a signal at smaller Euclidean time separations. To this end, we apply Gaussian smearing to the quark fields \cite{Gusken:1989qx,Alexandrou:1992ti},
\begin{equation}
  q^a(\vec{x},t) \rightarrow \tilde{q}^a(\vec{x}, t) = \sum_{\vec{y}} \l[\mathds{1} + \alpha_\mathrm{G} H^{ab}(\vec{x},\vec{y};U(t))\r]^{N_G} q^b(\vec{y}, t) \,, \quad q=u,d \,,
  \label{eq:Gaussian_smearing}
\end{equation}
with a choice of $\alpha_\mathrm{G}=0.2$ and $N_G=125$ smearing steps corresponding to a smearing radius of $r_G=0.47\fm$ in physical units \cite{Alexandrou:2019ali}. The hopping matrix $H^{ab}(\vec{x},\vec{y};U(t))$ is given by
\begin{equation}
 H^{ab}(\vec{x},\vec{y}; U(t)) = \sum_{i=1}^{3} \l[ U_i^{ab}(x)\delta_{x,y-\hat{i}} + U_i^{*ba}(x-\hat{i})\delta_{x,y+\hat{i}} \r] \,,
\end{equation}
where $a$, $b$ are color indices. Furthermore, we use APE-smeared \cite{Albanese:1987ds} gauge links $U$ in the construction of $H$ with a smearing parameter of $\alpha_\mathrm{APE}=0.5$ and $N_\mathrm{APE}=50$ smearing steps. \par

For the computation of three-point functions, we use sequential inversions through the sink \cite{Dolgov:2002zm}. Therefore, we need to perform separate inversions for each choice of the source-sink time separation, the projector index and---in principle---the momentum at sink. However, the latter is fixed to be zero, as mentioned before. Since we are interested in the nucleon charge radius, it is sufficient to consider three-point functions projected with $\Gamma_0$. Three-point functions are computed for five values of $\tsep$ as listed in Table~\ref{tab:tsep}. In order to obtain comparable effective statistics at each value of $\tsep$, we add additional source positions for increasing values of $\tsep$. The number of source positions $N_\mathrm{s}$ per source-sink time separation has also been included in Table~\ref{tab:tsep} together with the total number of measurements $N_\mathrm{meas}$. The source positions are randomly and independently chosen on each gauge configuration. Two-point functions are computed with matching statistics for each value of $\tsep$ from the forward propagators obtained in the calculation of three-point functions; hence, there is a possible choice of either using matching statistics for two- and three-point functions at each value of $\tsep$ or always using the full available statistics for the two-point functions. However, we found that fully preserving correlation by using exactly matching statistics yields a slight advantage with respect to the resulting statistical fluctuations. Calculations are performed using an appropriately tuned multigrid algorithm \cite{Bacchio:2016bwn,Bacchio:2017pcp,Alexandrou:2016izb} for the efficient inversion of the Dirac operator that is required for the computation of the quark-connected diagrams. \par

\begin{table}[!t]
 \centering
  \begin{tabular}{llll}
   \hline\hline
   $\tsep/a$ & $\tsep/\mathrm{fm}$ & $N_\mathrm{s}$ & $N_\mathrm{meas}$ \\
   \hline\hline
   12 & 0.97 & 4  & 3000  \\
   14 & 1.13 & 6  & 4500  \\
   16 & 1.29 & 16 & 12000 \\
   18 & 1.46 & 48 & 36000 \\
   20 & 1.62 & 64 & 48000 \\
   \hline\hline
  \end{tabular}
  \caption{Overview of the values of $\tsep$ used in the computation of three-point functions in lattice and physical units. $N_\mathrm{s}$ is the number of source positions for each source-sink time separation on each of the 750 gauge configuration and $N_\mathrm{meas}$ the corresponding, total number of measurements.}
 \label{tab:tsep}
\end{table}

\subsection{Ratio method}
Extracting the physical matrix elements in Eq.~(\ref{eq:matrix_element}) requires the cancellation of unknown overlap factors in the three-point function, which can be achieved by forming an optimized ratio involving two-point functions \cite{Alexandrou:2006ru,Alexandrou:2011db,Alexandrou:2013joa}
\begin{equation}
 R_{\mu}(\Gamma_\nu,\vec{q},t,\tsep) = \frac{C_\mu(\Gamma_\nu, \vec{q}, t, \tsep)}{\Ctwopt{\tsep}{0}} \sqrt{\frac{\Ctwopt{\tsep-t}{q}\Ctwopt{t}{0}\Ctwopt{\tsep}{0}}{\Ctwopt{\tsep-t}{0}\Ctwopt{t}{q}\Ctwopt{\tsep}{q}}} \,.
 \label{eq:ratio}
\end{equation}
At large Euclidean time separations $t$ and $t_f-t$, the ground state contribution is expected to dominate asymptotically and the ratio approaches a plateau,
\begin{equation}
 \lim_{t\rightarrow\infty} \lim_{t_f-t\rightarrow\infty} R_{\mu}(\Gamma_\nu,\vec{q},t,\tsep) = \Pi_\mu(\Gamma_\nu,\vec{q}) \,.
 \label{eq:plateau_method}
\end{equation}
The electromagnetic Sachs form factors can be extracted from $\Pi_\mu(\Gamma_\nu,\vec{q})$ for appropriate choices of insertion and projector indices,
\begin{align}
 \Pi_0\l(\Gamma_0,\vec{q}\r) &= -C\frac{E_N\l(\vec{q}\r)+m_N}{2m_N} G_E\l(Q^2\r) \,, \label{eq:G_E_temporal_insertion} \\
 \Pi_i\l(\Gamma_0,\vec{q}\r) &= -C\frac{i}{2m_N} q_i G_E\l(Q^2\r) \,, \label{eq:G_E_spatial_insertion} \\
 \Pi_i\l(\Gamma_k,\vec{q}\r) &= -C\frac{1}{4m_N} \epsilon_{ijk} q_j G_M\l(Q^2\r) \,, \label{eq:G_M_spatial_insertions} 
\end{align}
where $C=\sqrt{\frac{2m_N^2}{E_N\l(\vec{q}\r)\l(E_N\l(\vec{q}\r)+m_N\r)}}$. For the computation of the nucleon electric charge radius, we use the first relation, which gives by far the best signal for $G_E(Q^2)$ and in addition allows to obtain a result directly at zero momentum transfer. 

The determination from the second relation involving $G_E(Q^2)$ is impeded due to the momentum factor appearing on the right-hand side 
of the equation and similarly in the last equation for $G_M(Q^2)$. A direct method for derivatives of form factors
at zero momentum has been put forward in \cite{deDivitiis:2012vs}, based on the algebraic definition of the momentum derivative and quark propagator
expansion, which promotes the n-point correlation function by one point per derivative and is thus very costly.

In Ref.~\cite{Alexandrou:2016rbj} we have explored model-independent position space methods to remedy this issue for $G_M(Q^2)$ without resorting to fits. We remark that one of these methods called momentum elimination in the plateau-region is similar to the approach we will introduce in the next section to allow for a direct, model-independent computation of the nucleon electric charge radius, which is otherwise hindered by the discrete nature of momenta on the lattice. \par

\subsection{Summation method}
In nucleon structure calculations it is notoriously difficult to reach ground state dominance due to an exponential signal-to-noise problem; hence, a careful analysis of the corresponding systematics is required. While we have lattice data available for five values of $\tsep$ as listed in Table~\ref{tab:tsep}, it is not \emph{a priori} clear that this is sufficient to control excited state effects. Therefore, we use the so-called summation method \cite{Maiani:1987by,Dong:1997xr,Capitani:2012gj} in addition to the direct plateau method, which allows for a stronger suppression of excited states 
\begin{equation}
 \sum\limits_{t=t_\mathrm{ex}}^{t_\mathrm{sep}-t_\mathrm{ex}} R_\mu(\Gamma_\nu, \vec{q}, t, \tsep) = \mathrm{const} + (\tsep - 2t_\mathrm{ex} + a) \cdot \Pi_\mu\l(\Gamma_\nu,\vec{q}\r) + \mathcal{O}(e^{-\Delta \tsep}) \,,
 \label{eq:summation_method}
\end{equation}
where $t_\mathrm{ex}=2a$ for the conserved vector current insertion. The contribution from the next higher state with a mass gap $\Delta$ is now suppressed by an additional factor $e^{-\Delta \tsep}$ in contrast to the plateau method for which the suppression is only $\sim e^{-\Delta t}$. \par

\section{Position space method for \texorpdfstring{$\langle r^2_E \rangle$}{the nucleon radius}} \label{sec:method}
The mean-square charge radius of the nucleon $\langle r_E^2\rangle$ is defined from the electric Sachs form factor in Eq.~(\ref{eq:G_E}) through expansion around small $Q^2$,
\begin{equation}
 G_E(Q^2) = G_E(0) \l[ 1 - Q^2 \langle r_E^2\rangle/6 + \mathcal{O}(Q^4) \r] \,.
 \label{eq:G_E_expansion}
\end{equation}
Since for the proton and the isovector combination one has $G_E^{p,u-d}(0)=1$, the mean-square charge radius can be extracted from the expansion via computing
\begin{equation}
 \langle r_E^2 \rangle = \l. -6 \frac{dG_E(Q^2)}{dQ^2}\r|_{Q^2=0} \,.
 \label{eq:ms_radius}
\end{equation}
For the neutron, the leading term in Eq.~(\ref{eq:G_E_expansion}) is absent and the normalization factor $G_E^n(0)=0$ is dropped in the definition to obtain a finite result; hence, $\langle r_E^2 \rangle^n$ can be computed in the same way from Eq.~(\ref{eq:ms_radius}). Moreover, for the proton and the isovector combination, it is possible to define a root mean-square charge radius, i.e.,
\begin{equation}
 r_E^{p,u-d} = \sqrt{\langle r_E^2 \rangle^{p,u-d}} \,,
 \label{eq:rms_radius}
\end{equation}
which is not meaningful for the neutron as its mean-square charge radius $\langle r_E^2 \rangle^n$ is negative. \par

Computing $\langle r_E^2 \rangle$ requires an evaluation of the derivative with respect to $Q^2$ which is not directly possible in a finite box as only finite, discrete momenta are accessible. In the literature, several methods have been used to circumvent this issue, e.g., dipole fits. However, any such method introduces model dependence and potentially large, systematic uncertainties considering the physical values of momenta that are typically available in lattice simulations and the steepness of $G_E(Q^2)$ close to zero momentum transfer. For a recent review on common methods to parametrize electromagnetic form factors, we refer to Ref.~\cite{Punjabi:2015bba}. \par

In this study, we aim to evaluate the pertinent derivative in Eq.~(\ref{eq:ms_radius}) by an integral method for a suitably defined form factor in the small momentum region given below in Eqs.~(\ref{eq:Pi_of_khat_2}) and (\ref{eq:Pi_of_khat_3}). It allows for systematic probing of the dependence on the available supporting lattice data and has a well-defined infinite volume as well as continuum limit without any further adjustments of parametrization. \par

We present the calculation in its simplest one-dimensional form based on (numerical) form factor data for on-axis three-momenta ( $\qvec \propto \evec_k$). It can be extended to incorporate arbitrary momentum directions, which, however, require taking into account anisotropy effects and analytic continuation. \par

In the continuum we have $Q^2 = -2 m_N\,\left( E_N (q) - m_N \right)$ given our momentum setup for the nucleon at source and sink, and we can straightforwardly replace the $Q^2$ derivative by 
\begin{align}
  \frac{d}{d Q^2} &= -\frac{E_N(q)}{m_N}\,\frac{d}{d q^2}
  \label{eq:chain_rule}
\end{align}
for on shell $E_N = \sqrt{m_N^2 + q^2}$. In particular, we can use
\begin{align}
  \left. \frac{d}{dQ^2} \,G_E(Q^2) \right|_{Q^2 = 0} &= -\lim\limits_{q \to 0}\, \frac{G_E(q) - G_E(0)}{q^2} \,.
  \label{eq:derivative_limit}
\end{align}
Thus, on the lattice, we define the form factor
\begin{align}
 \Pi( q ) = -\frac{1}{6}\, \sqrt{\frac{2 m_N E_N(q)}{E_N(q) + m_N}} \,   \, \sum\limits_{\sigma = \pm, \, k = 1,2,3} \,
 \Pi_0 \left(\Gamma_0, \qvec = \sigma \, q \, \evec_k \right) \,,\quad q = 2\pi/L \times \left\{ 0,1,2,\ldots, \qmax \right\} \,.
 \label{eq:Pi_of_q}
\end{align}
In Eq.~(\ref{eq:Pi_of_q}), $\qmax$ denotes the maximal value of on-axis momentum, for which numerical data for the ratio $R_\mu$ and thus the form factor $\Pi_0$ can be obtained given the achieved statistical precision for two- and three-point correlation functions. Apart from the dependence on the source-sink time separation $\tsep$, checking the saturation of the integral defining $r_E^2$ under variation of $\qmax$ will be a major point of the systematic error analysis. To this end, we will also use model data to study the influence of the large-$Q^2$ tail, i.e. allowing us to implement larger values of $\qmax$ than the statistical precision of the available lattice data would otherwise permit.

\begin{figure}[t]
 \centering
 \subfigure{\includegraphics[totalheight=0.225\textheight]{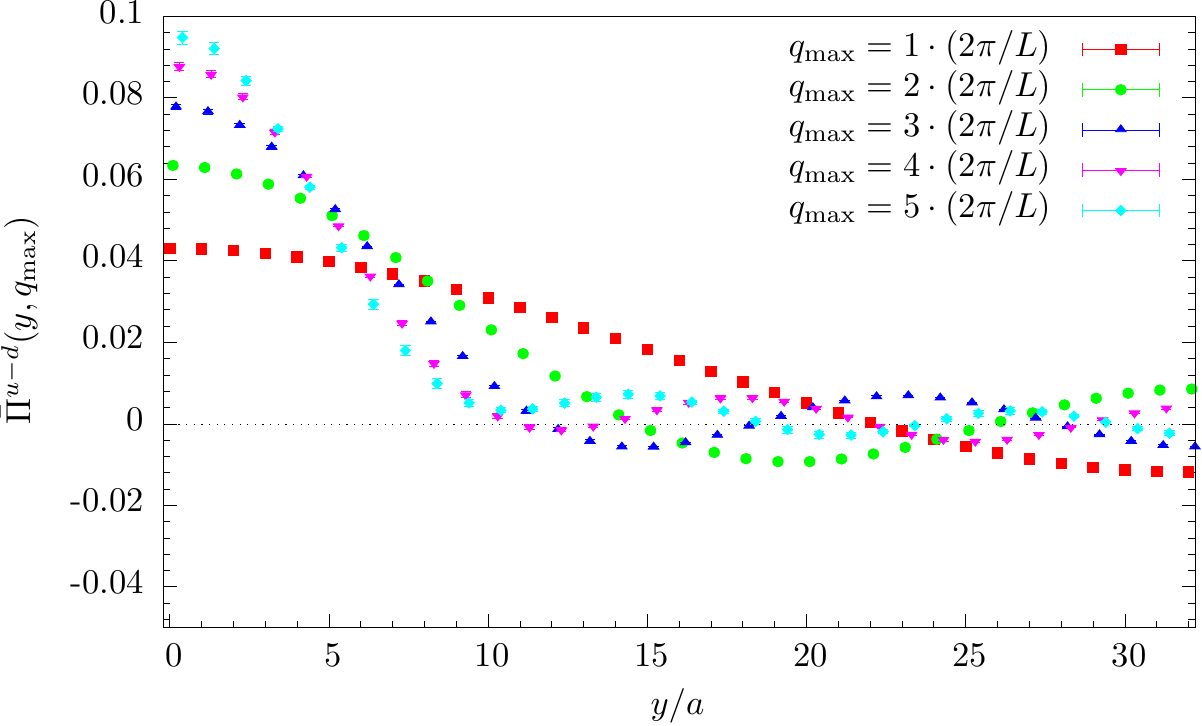}} \quad
 \subfigure{\includegraphics[totalheight=0.225\textheight]{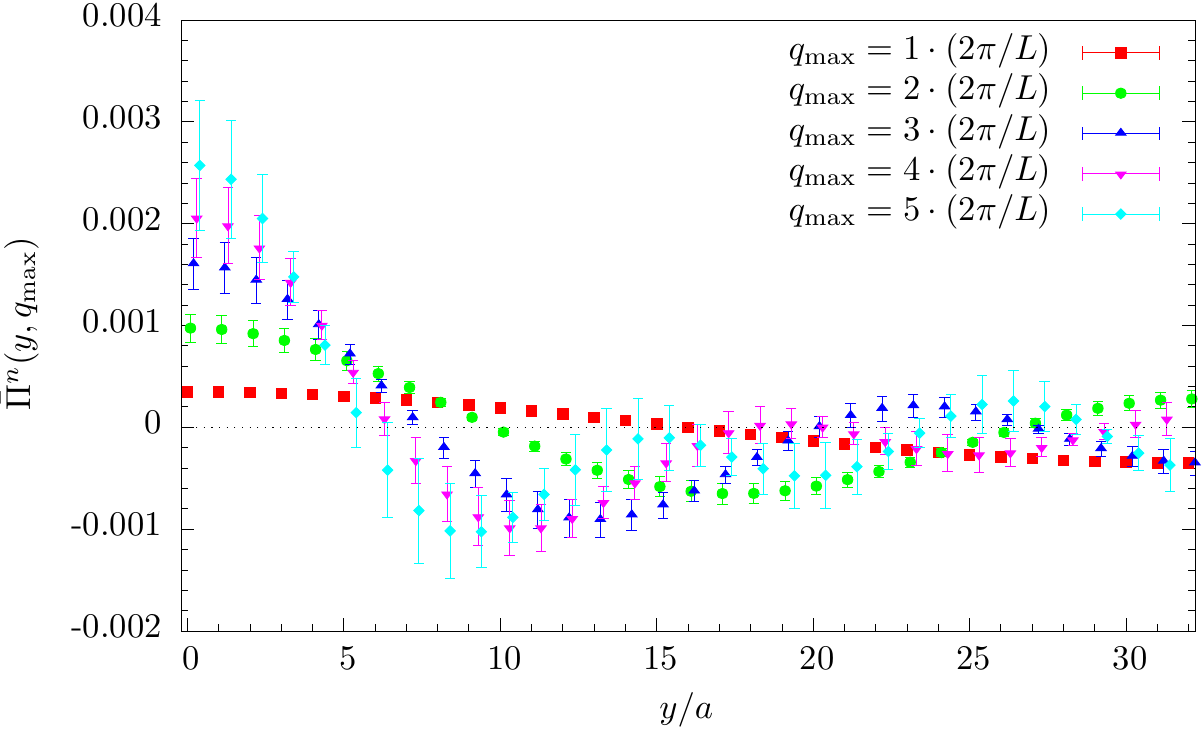}}
 \caption{Position space lattice data $\Pi(y)$ for different choices of the momentum cutoffs $\qmax$ and $\tsep=0.97\fm$. Results are shown for the isovector combination (left panel) and the neutron (right panel). Data for different values of $\qmax$  have been displaced horizontally to improve readability.}
 \label{fig:Pi_y}
\end{figure}

Together with $\Pi(q)$ in Eq.~(\ref{eq:Pi_of_q}) we obtain its counterpart in position space $\Pibar(n)$ from the discrete Fourier transform for $-N/2 \le n \le N/2$ with $n=y/a$, $N=L/a$ and $\Pibar(-y) = \Pibar(y)$. Sample data for the position space form factor $\Pibar(y,\qmax)$ with varying cutoff $\qmax$ in the discrete Fourier transform are shown in Fig.~\ref{fig:Pi_y}.

\begin{figure}[htpb]
  \centering
  \includegraphics[width=0.495\textwidth]{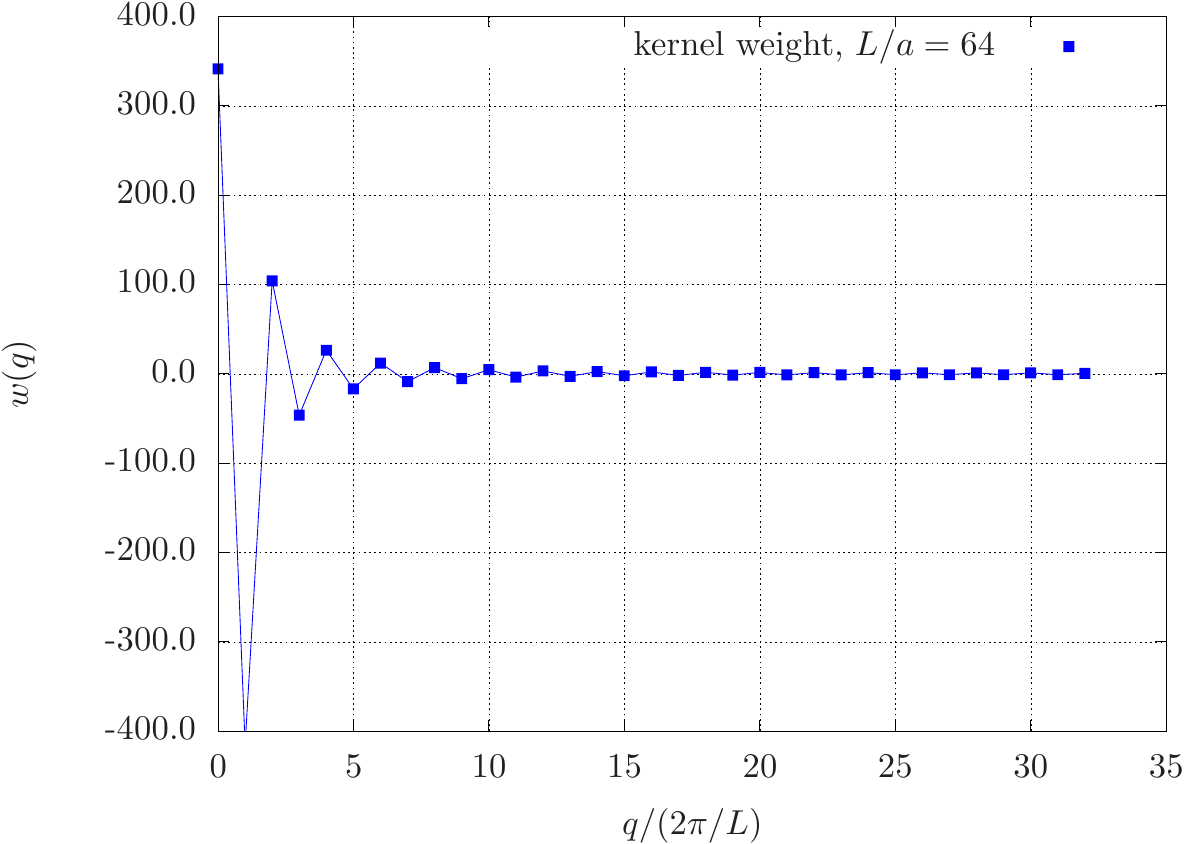}
  \includegraphics[width=0.495\textwidth]{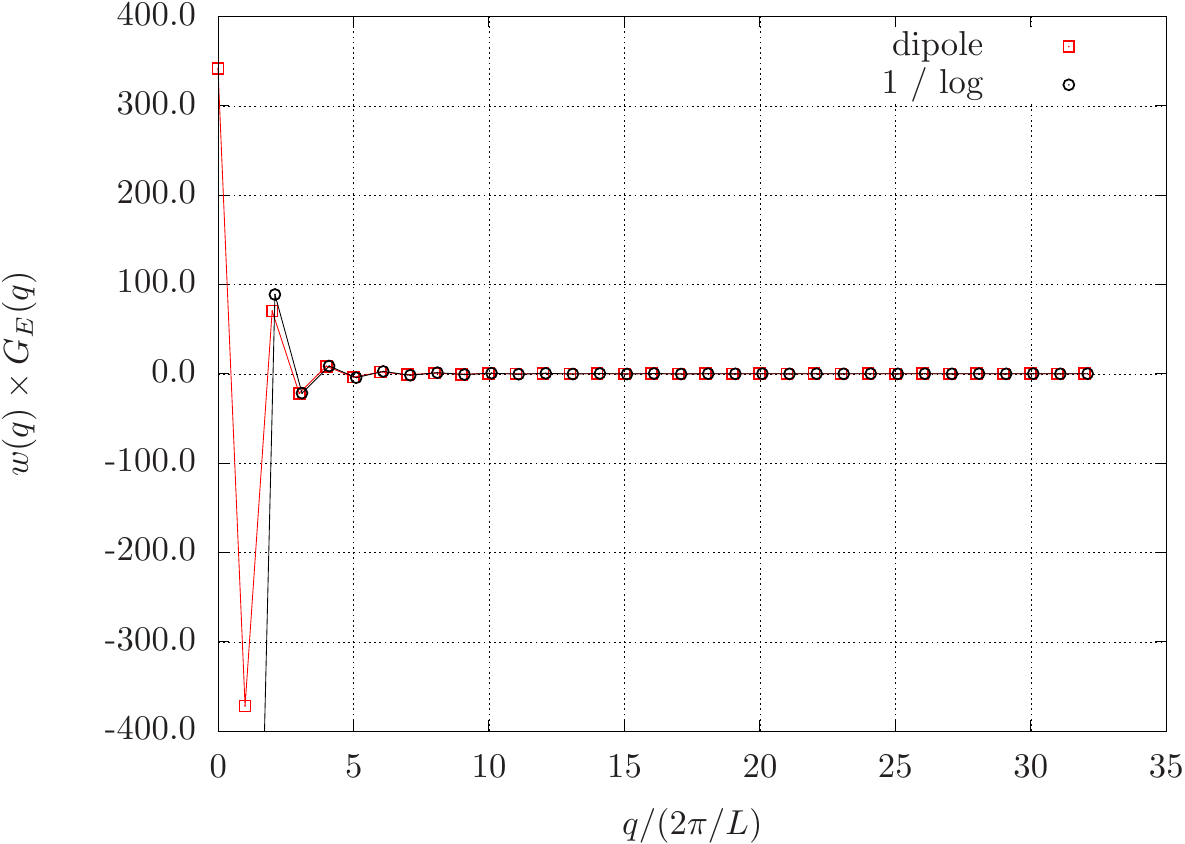}
  \caption{Left: kernel weight $w(q)$ of Eq.~(\ref{eq:Pi_of_khat_7}); right: integrand (kernel weight times evaluated fit for $G_E(q)$) for dipole- (red open squares) and inverse logarithmic-type decay in the large momentum tail region ($q / (2\pi/L)\gtrsim 7$). The continuous lines are added for visual guidance.}
  \label{fig:khat2_eq_0_kernel_weight}
\end{figure}

Upon inverse Fourier transform, we can express the resulting form factor as a function of the lattice momentum $\khat = 2\,\sin(k/2)$ via
\begin{align}
  \Pi\left( \khat^2 \right) &= \sum\limits_{n=0}^{N/2} \; c_n \, \Pibar(n)\;\cos\left( 2n \, \asin\left( \khat/2 \right) \right)
  \label{eq:Pi_of_khat} \\
  c_n &= \left\{
    \begin{matrix}
      2 & n = 1,\ldots, N/2-1 \\
      1 & n = 0,\, N/2 \,.
    \end{matrix}
    \right.
    \nonumber
\end{align}
Using the relation
\begin{align}
  \cos\left( 2n\,\asin(\khat/2)  \right) &=
  (-1)^n\,T_{2n}\left( \khat/2 \right)
  \label{eq:chebyshev_1}
\end{align}
where $T_{2n}$ are Chebyshev polynomials of the first kind, this gives
\begin{align}
  \Pi\left( \khat^2 \right) &= \sum\limits_{n=0}^{N/2} \; c_n \, \Pibar(n)\;(-1)^n\,T_{2n}(\khat/2)
  \label{eq:Pi_of_khat_2}
\end{align}
and with respect to Eq.~(\ref{eq:derivative_limit})
\begin{align}
  \frac{1}{\khat^2}\; \left( \Pi( \khat^2) - \Pi( 0 ) \right) &= 
  \sum\limits_{n=1}^{N/2} \; c_n \, \Pibar(n)\; P_n(\khat^2)
  \label{eq:Pi_of_khat_3}
\end{align}
where
$P_n(\khat^2) = \left[ (-1)^n T_{2n}(\khat/2) - 1 \right] / \khat^2$ are polynomials in $\khat^2$ of degree $n-1$. The difference quotient in Eq.~(\ref{eq:Pi_of_khat_3}) is an
analytic function with a well-defined infinite volume and continuum limit and gives in particular
\begin{align}
  \brackets{r_E^2} &= \lim\limits_{\khat^2 \to 0} \, \frac{1}{\khat^2}\; \left( \Pi( \khat^2) - \Pi( 0 ) \right) 
  = \sum\limits_{n=1}^{N/2} \; c_n \, \Pibar(n)\; P_n(0) \,.
  \label{eq:Pi_of_khat_4}
\end{align}
Since for the Chebyshev polynomials $T_{2n}(x) = (-1)^{n+1}\,\frac{(2n)^2}{2}\,x^2 + \order{x^4}$, we also have the familiar integral / summation form
\begin{align}
  \brackets{r_E^2} &\propto  \sum\limits_{n=1}^{N/2} \; c_n \, \Pibar(n)\;(2n)^2 \,,
  \label{eq:Pi_of_khat_5}
\end{align}
or equivalently in terms of the original data $\Pi(q)$ in Eq.~(\ref{eq:Pi_of_q}),
\begin{align}
  \brackets{r_E^2} &\propto  \sum\limits_{q=0}^{\qmax} \, w(q)\, \Pi(q) 
  \label{eq:Pi_of_khat_6} \\
  w(q) &= \sum\limits_{n=1}^{N/2} \; c_n \;(2n)^2 \, \cos(q n)  \,.
  \label{eq:Pi_of_khat_7}
\end{align}
The kernel weight $w$ in Eq.~(\ref{eq:Pi_of_khat_6}) for the lattice setup used in this study and exemplary data for the integrand in Eq.~(\ref{eq:Pi_of_khat_7}) are shown in Fig.~\ref{fig:khat2_eq_0_kernel_weight}.
Inspection of the integrands in the right-hand panel of Fig.~\ref{fig:khat2_eq_0_kernel_weight} shows that the contribution from the form factor at on-axis momenta beyond 
$q /(2\pi/L) = 5 \sim 7$ will not be significant given our currently achieved statistical accuracy of the form factor.

\section{Results} \label{sec:results}
\begin{figure}[htp]
 \centering
 \subfigure{\includegraphics[totalheight=0.3\textheight]{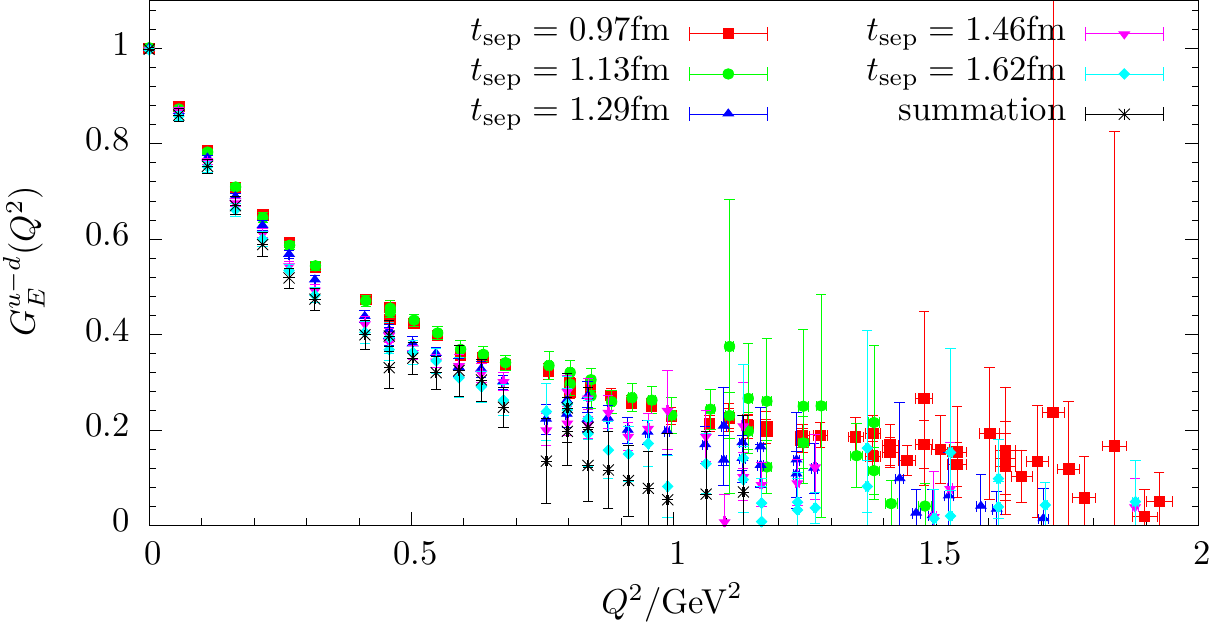}}
 \subfigure{\includegraphics[totalheight=0.3\textheight]{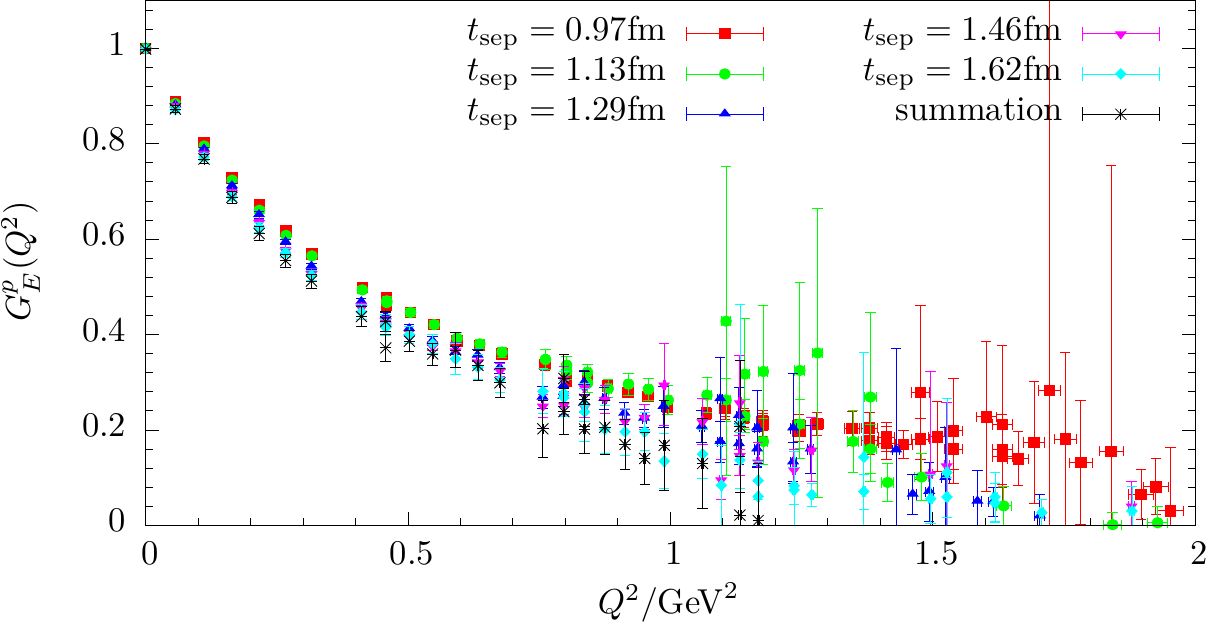}} \\
 \subfigure{\includegraphics[totalheight=0.3\textheight]{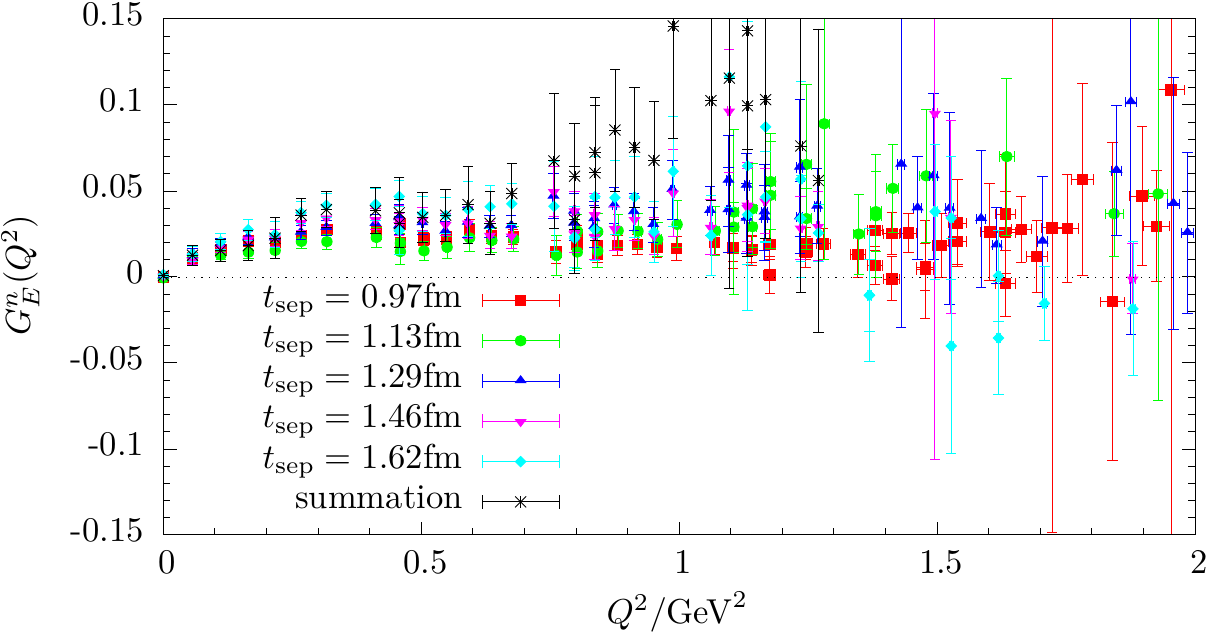}}
 \caption{Lattice data for the ratio in Eq.~(\ref{eq:ratio}) for $G_E(Q^2)$ at all five source-sink time separations for all three isospin combinations of the operator insertion. Only quark-connected contributions are included.}
 \label{fig:G_E_vs_Qsqr}
\end{figure}

The lattice data for the electric Sachs form factor that are used as input in our calculation of the nucleon radii are shown in Fig.~\ref{fig:G_E_vs_Qsqr} as a function of $Q^2$. We have included data from the ratio method in Eq.~(\ref{eq:plateau_method}) at all five available source-sink time separations as well as from the summation in Eq.~(\ref{eq:summation_method}). Results are shown for the proton, neutron and the isovector combination. In any case, we observe that the summation method yields results compatible with the ratio method for the largest value of $\tsep$.\par

\begin{figure}[ht!]
 \centering
  \subfigure{\includegraphics[totalheight=0.2\textheight]{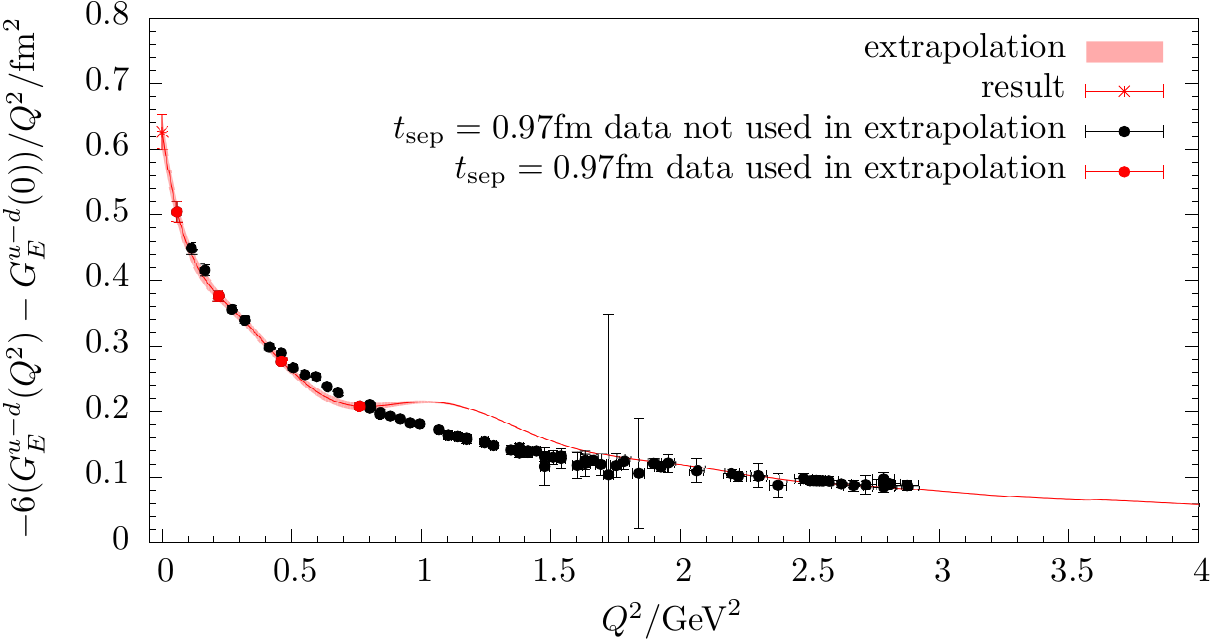}}
  \subfigure{\includegraphics[totalheight=0.2\textheight]{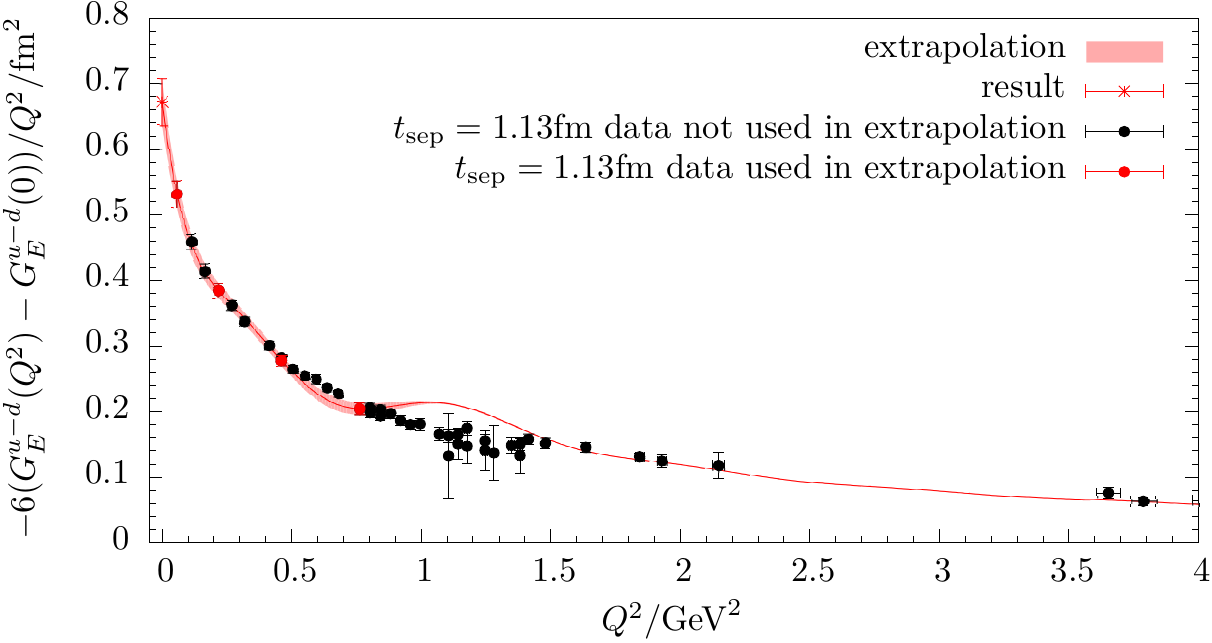}} \\
  \subfigure{\includegraphics[totalheight=0.2\textheight]{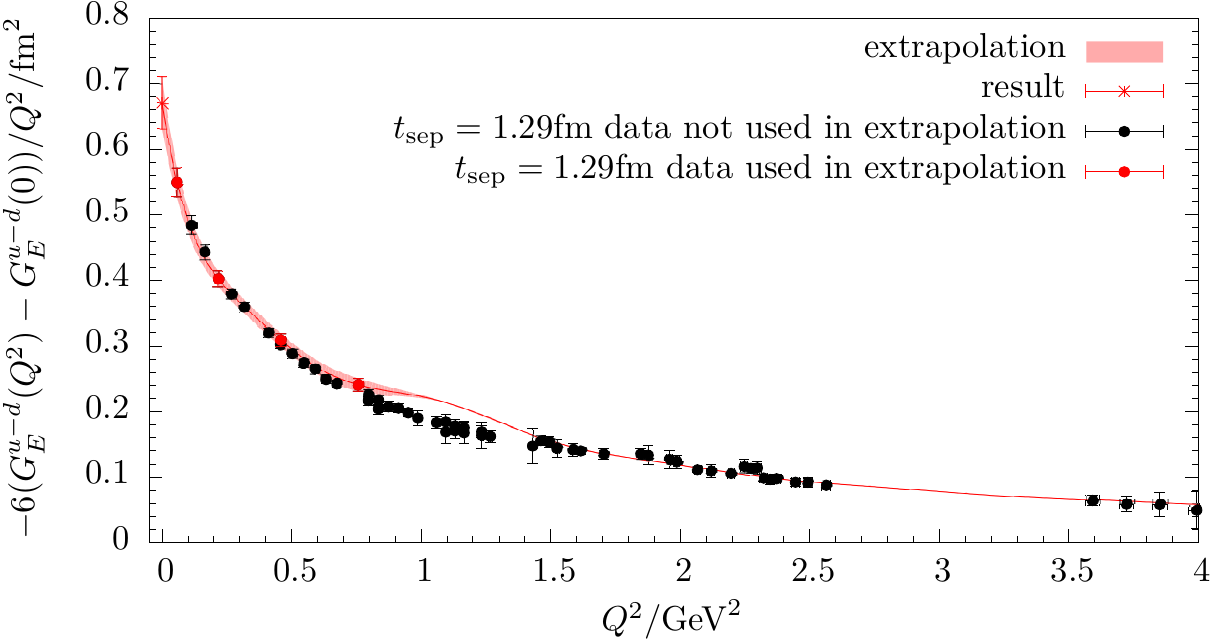}}
  \subfigure{\includegraphics[totalheight=0.2\textheight]{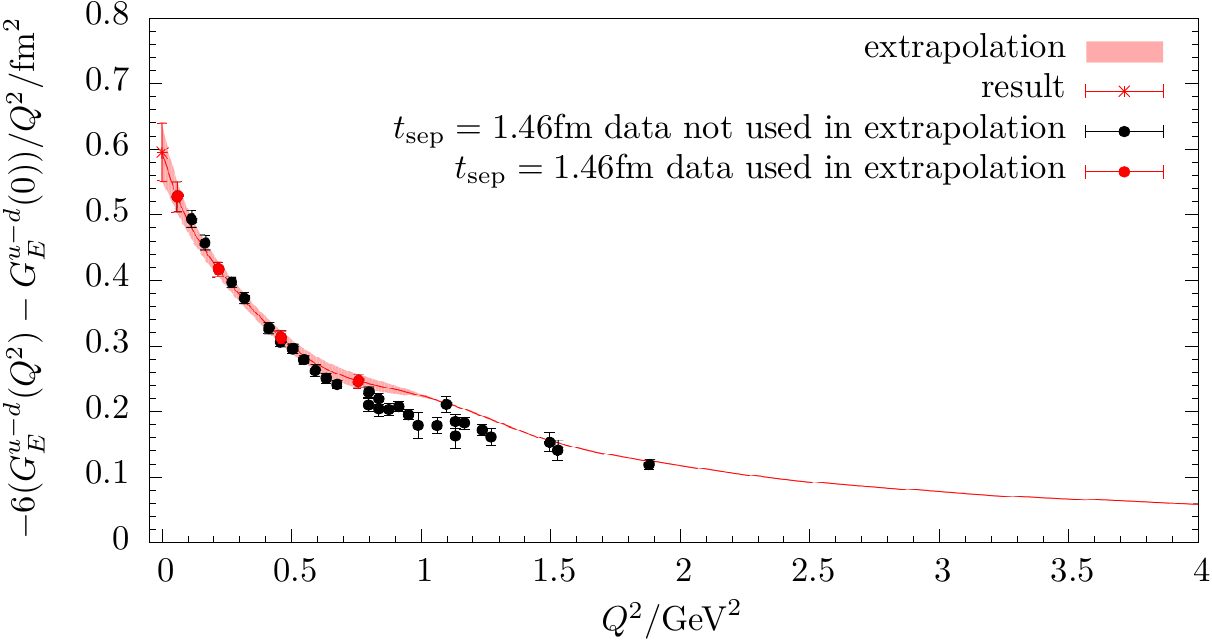}} \\
  \subfigure{\includegraphics[totalheight=0.2\textheight]{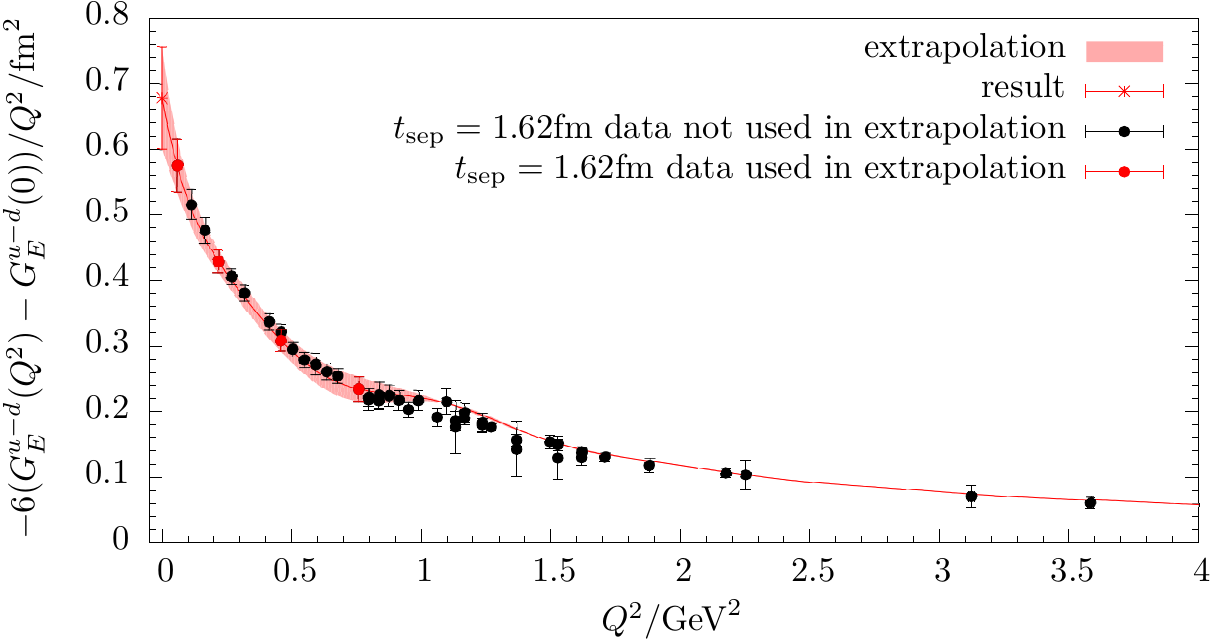}}
  \subfigure{\includegraphics[totalheight=0.2\textheight]{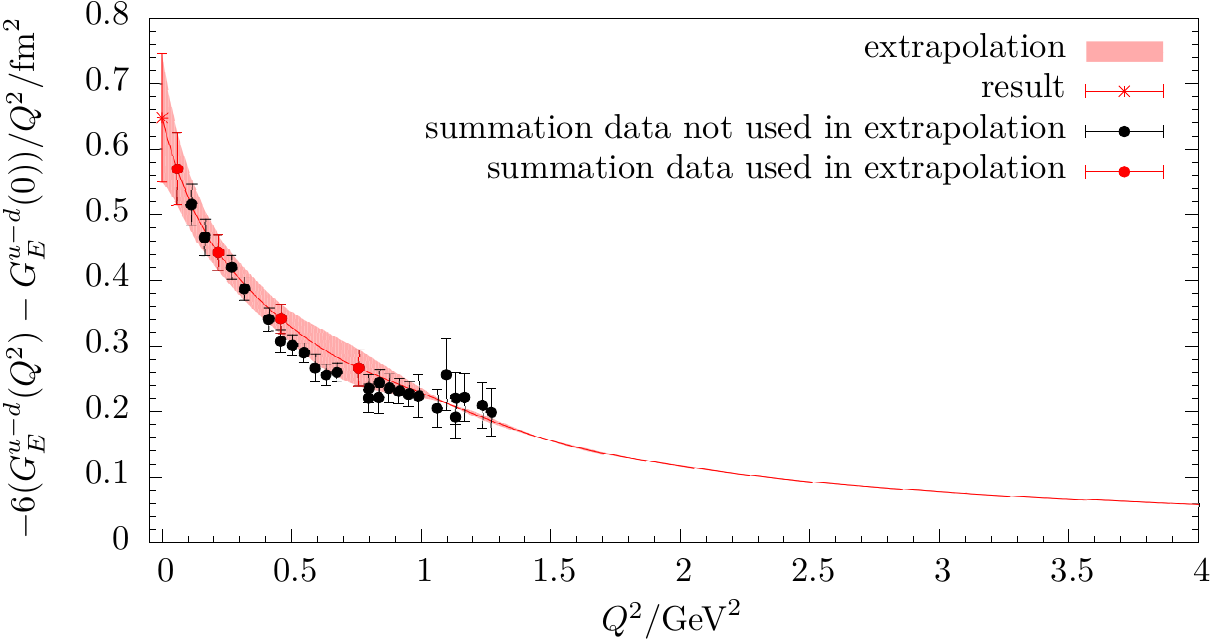}}
 \caption{Lattice data for $-6 (G_E(Q^2)^{u-d}-G_E^{u-d}(0))/Q^2$ and continuous band computed from Eq.~(\ref{eq:master_formula}) using $\qmax=4\cdot 2\pi/L$ for all five source-sink time separations and the summation method.}
 \label{fig:iso_rsqr_all_ts_summation}
\end{figure}

In Fig.~\ref{fig:iso_rsqr_all_ts_summation}, we show our lattice data for the difference $-6(G_E^{u-d}(Q^2)-G_E^{u-d}(0))/Q^2$ as a function of $Q^2$ in physical units for all five values of $\tsep/a$ and the summation method. The extrapolation bands are computed by evaluating Eq.~(\ref{eq:Pi_of_khat_3}) for any given value of
\begin{equation}
 \hat{k}^2(Q^2) = 4\sin^2\l(\frac{1}{2}\sqrt{\frac{Q^4}{4m^2} + Q^2}\r)
 \label{eq:khat_Qsqr}
\end{equation}
and multiplying the expression by $\hat{k}^2/Q^2$ leading to
\begin{equation}
 \frac{\Pi(Q^2)-\Pi(0)}{Q^2} = -\sum_{n=1}^{N/2} c_n \bar{\Pi}(n) \frac{2 \sin^2\l( \frac{n}{2} \sqrt{\frac{Q^4}{4m^2}+Q^2} \r)}{Q^2} \,,
 \label{eq:master_formula}
\end{equation}
where $\sqrt{\frac{Q^4}{4m^2} + Q^2}=q$. Note that the factor related to substituting $\hat{k}^2\rightarrow Q^2$ is one for the nucleon radius, i.e., at $Q^2=\hat{k}^2=0$. However, at nonzero $Q^2$, the change of variable $\hat{k}^2\rightarrow Q^2$ has to be taken into account explicitly to make contact with the discrete lattice data, which is achieved by the above expression. Here we have consistently chosen a rather small value of $\qmax=4\cdot (2\pi/L)$ for the extrapolations in all six plots. Still, the resulting bands describe well the lattice data for $Q^2\lesssim0.5\gev$. Data points entering the construction of the extrapolation are shown in red and fall onto the curve by definition. At larger momenta we observe deviations from the lattice data, namely, a wiggling behavior in the band around $Q^2=1\gev^2$. This is expected as the values of $q$ entering the initial Fourier transform from momentum to position space are restricted to values of $Q^2$ below $0.8\gev^2$ for $\qmax=4\cdot (2\pi/L)$. The behavior is particularly visible for the data at smaller values of $\tsep$ which are statistically more precise. \par

\subsection{Study of the large-\texorpdfstring{$Q^2$}{Qsqr} tail contribution}\label{subsec:large_Qsqr_tail}
In order to further study the systematics associated with the truncation of the Fourier transform for $\Pi(q)$ we have explored ways of generating synthetic data for the tail of the form factor to avoid this truncation altogether. To this end, we use several fit models that are applied to the lattice data at low values of $Q^2$. In a second, step we use the jackknife samples for the fit parameters to generate synthetic data samples for values of $Q^2$ corresponding to on-axis momenta $\vec{q}=(q,0,0)^T$ with $q>\qmax$. This synthetic data are then used to perform the Fourier transform of $\Pi(q)$ for all $q>\qmax$ in addition to the actual lattice data at $q\leq\qmax$. A simple choice for the proton and isovector form factor is the dipole fit model
\begin{equation}
 G_E^{p,u-d}(Q^2) = \frac{A}{\l(1+BQ^2\r)^2} \,,
 \label{eq:dipole_fit}
\end{equation}
where $A$, $B$ are free parameters of the fit. For the neutron, we use a Galster-like parametrization~\cite{Galster:1971kv,Kelly:2004hm} instead of the dipole fit model. This model takes into account that $G_E^n(0)=0$,
\begin{equation}
 G_E^n(Q^2) = \frac{C Q^2}{\l(1 + DQ^2\r)\l(1 + \frac{Q^2}{\Lambda^2} \r)^2} \,.
 \label{eq:galster_fit}
\end{equation}
The free parameters of this model are denoted by $C$ and $D$, while $\Lambda^2=0.71\gev^2$ is treated as a constant. Since the dipole fit model approaches zero quadratically for increasing values of $Q^2$, one might anticipate that using synthetic data from this model yields a result not very different from the one obtained by a simple truncation in the Fourier transform of $\Pi(q)$ at a reasonable value of $\qmax$. Therefore, we consider a $1/\log$ fit model with an unphysically slower decrease at large values of $Q^2$,
\begin{equation}
 G_E(Q^2) = \frac{E}{\log\l(1+FQ^2\r)}
 \label{eq:log_fit}
\end{equation}
with two fit parameters $E$ and $F$, which allows us to further investigate the dependence on the choice of the model. This model can be applied to all three isospin combinations, but it requires a careful choice of the lower bound of the fit range as the model is divergent for $Q^2\rightarrow0$. \par

\begin{figure}[ht!]
 \centering
  \subfigure{\includegraphics[totalheight=0.195\textheight]{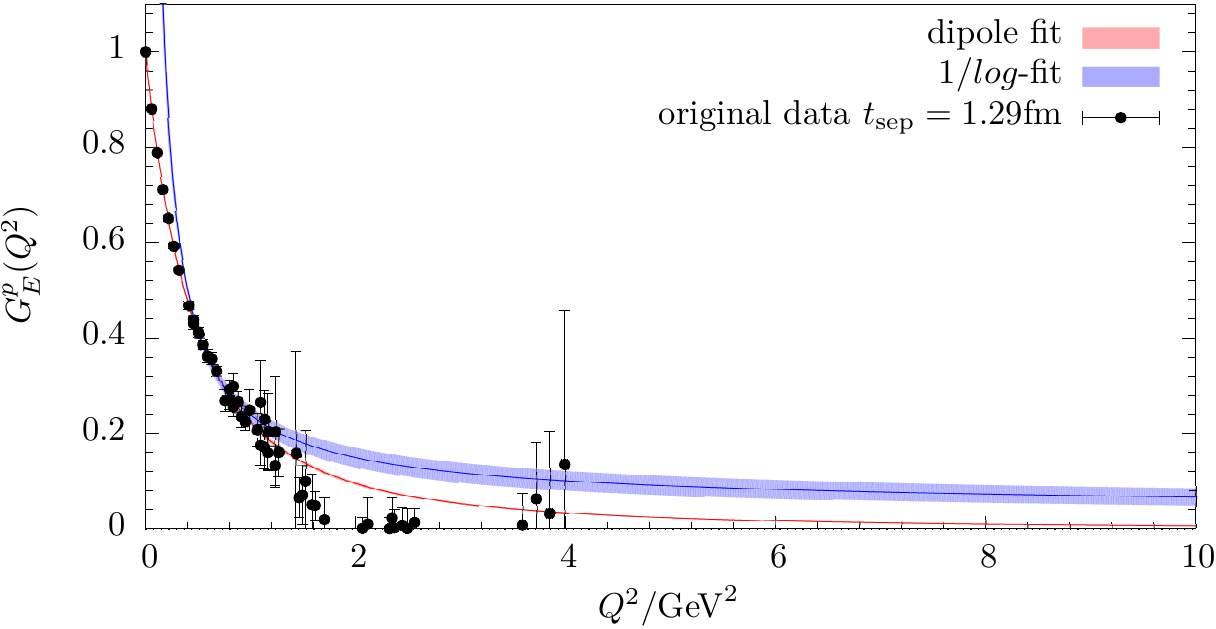}}
  \subfigure{\includegraphics[totalheight=0.200\textheight]{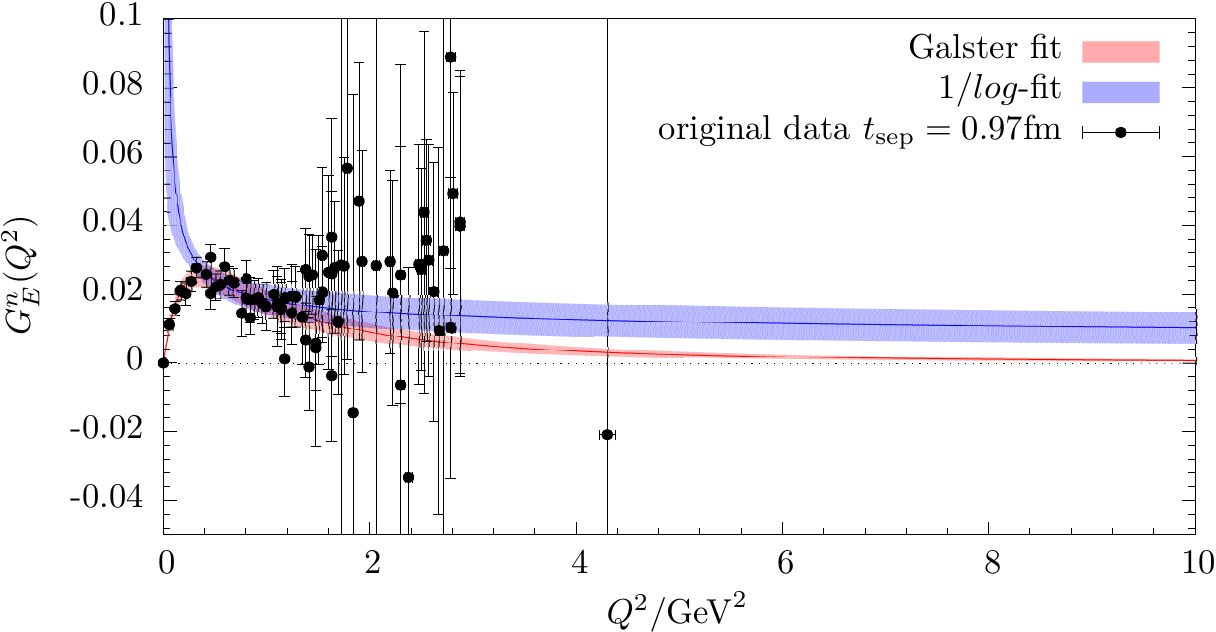}} 
 \caption{Different fit models of the lattice data for $G_E(Q^2)$. Left panel: resulting fit bands for the proton at $\tsep\approx1.29\fm$ from a dipole- and $1/\log$-fit. Right panel: resulting fit bands for the neutron at $\tsep\approx0.97\fm$ from a Galster-like fit and a $1/\log$-fit.}
 \label{fig:G_E_fits}
\end{figure}

In Fig.~\ref{fig:G_E_fits} results are shown for fitting the different models to the proton and neutron data for $G_E(Q^2)$. Corresponding results for the isovector insertion are very similar to the ones for the proton. As expected, the bands from the dipole and Galster-like fits exhibit a much sharper dropoff in the large-$Q^2$ tail of the form factor than the corresponding bands from the $1/\log$-fit. In general, fit ranges have been chosen such that $Q^2\in\l[0\gev^2,1.8\gev^2\r]$ for the dipole and Galster-like fits and $Q^2\in\l[0.7\gev^2,1.8\gev^2\r]$ for the $1/\log$-fit. We found this choice to yield good values of $\chi^2/\mathrm{d.o.f.}$ (degrees of freedom) regardless of the source-sink time separation as well as for data from the summation method. \par

\begin{figure}[ht!]
 \centering
  \subfigure{\includegraphics[totalheight=0.225\textheight]{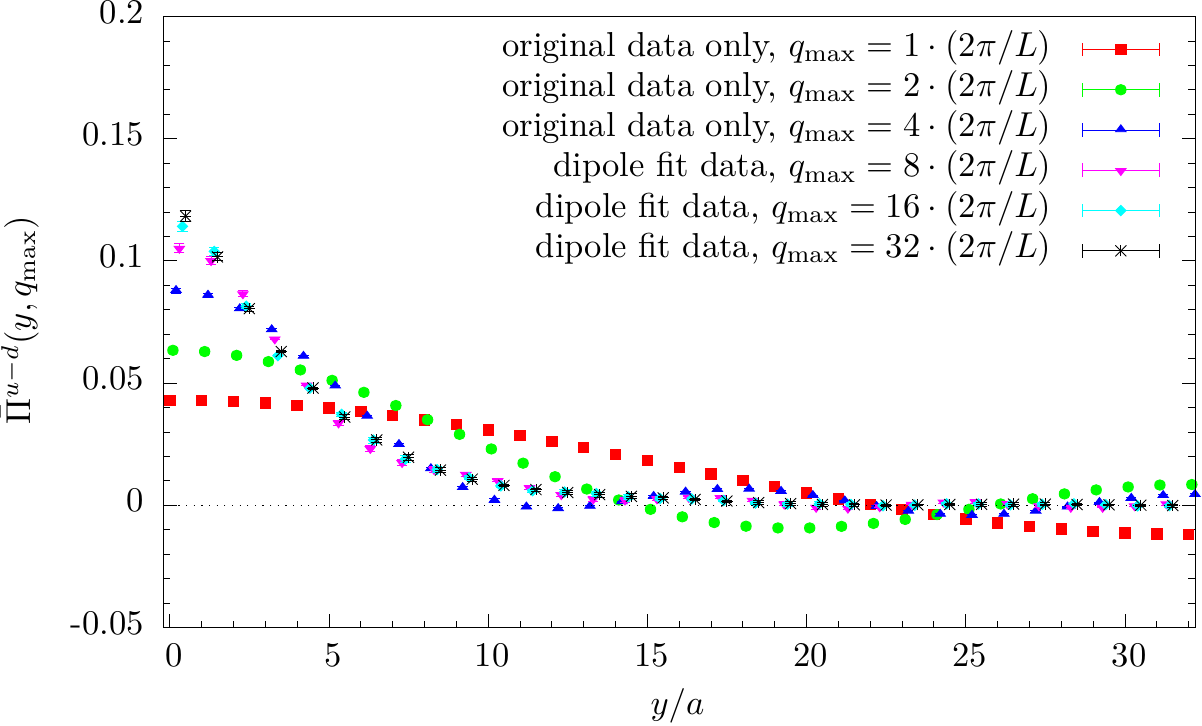}} \quad
  \subfigure{\includegraphics[totalheight=0.225\textheight]{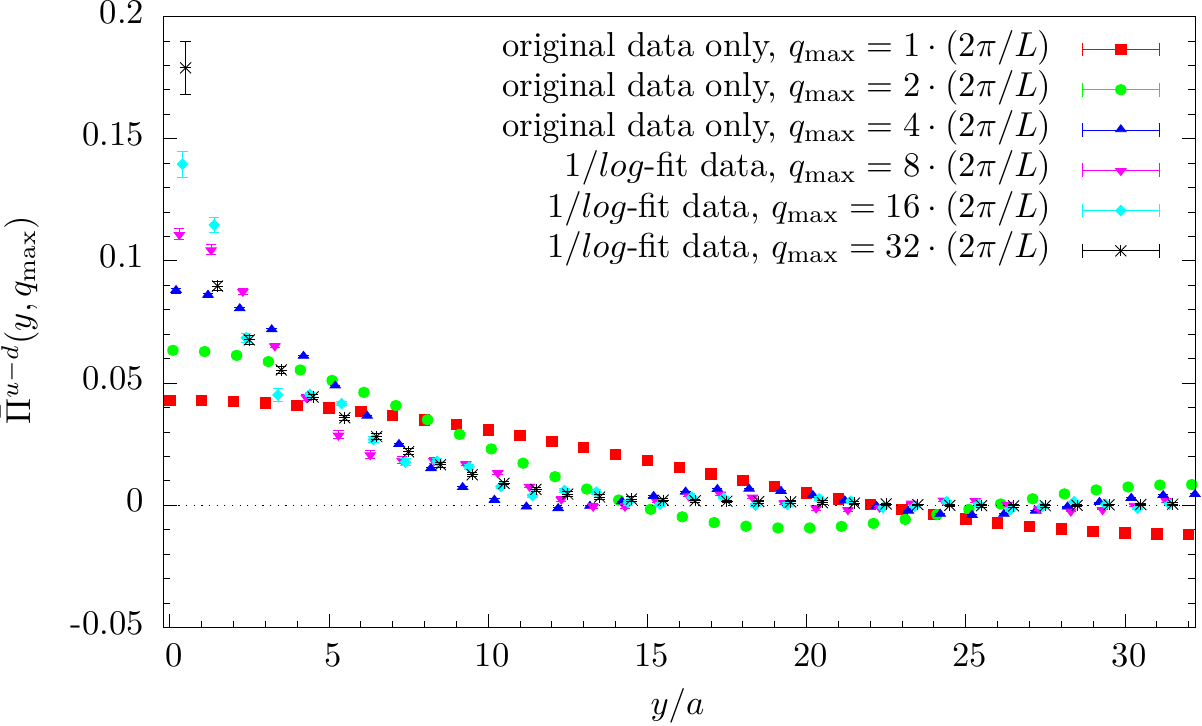}} \\ 
  \subfigure{\includegraphics[totalheight=0.225\textheight]{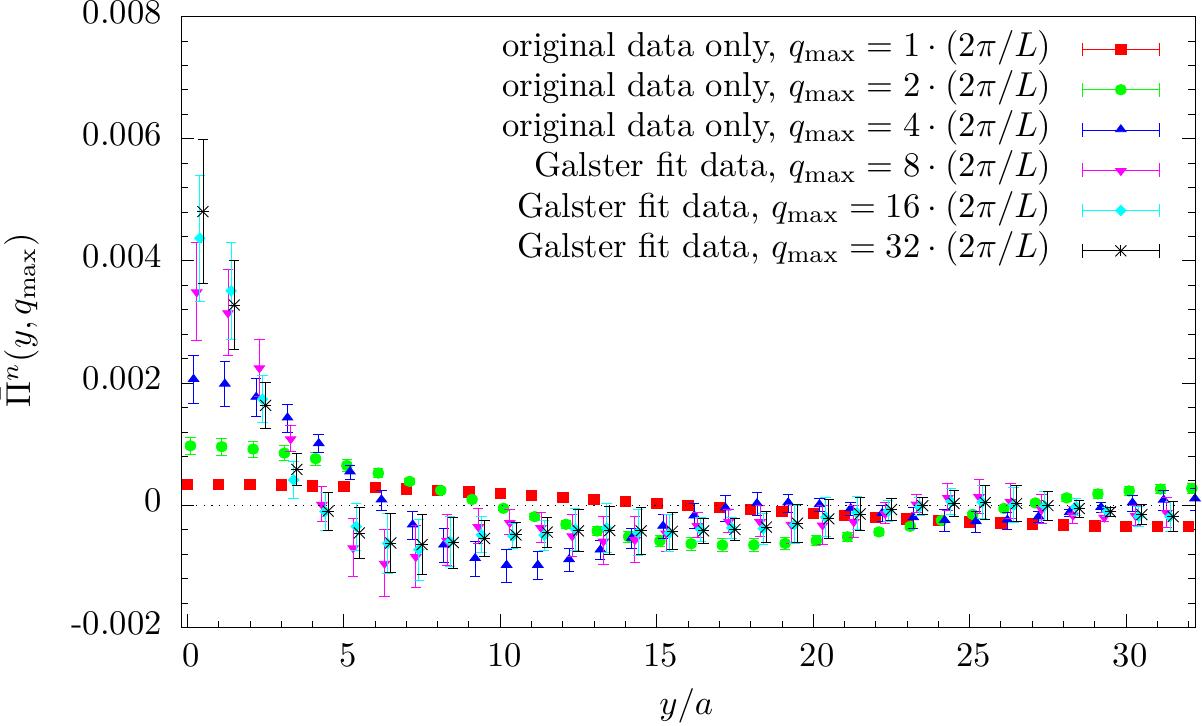}} \quad
  \subfigure{\includegraphics[totalheight=0.225\textheight]{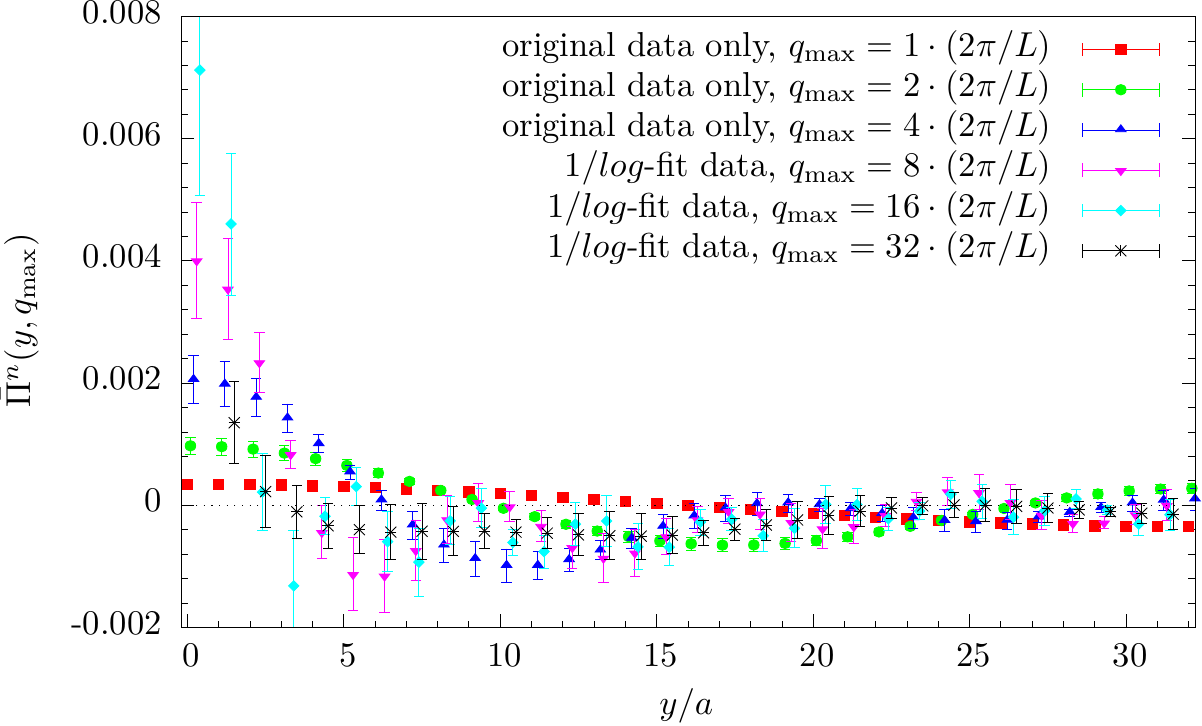}}
 \caption{Position space data $\bar{\Pi}(y,\qmax)$ for the isovector combination (upper row) and the neutron (lower row) for different choices of the momentum cutoff $\qmax$ and $\tsep=0.97\fm$. Results shown for $\qmax=8\cdot(2\pi/L)$ (pink), $\qmax=16\cdot(2\pi/L)$ (light blue) and $\qmax=32\cdot(2\pi/L)$ (black) use synthetic data generated from the fitted parameters of the models in Eqs.~(\ref{eq:dipole_fit})--(\ref{eq:log_fit}) in addition to the lattice data for $q\leq4\cdot(2\pi/L)$. Data for different values of $\qmax$ have been displaced horizontally to improve readability.}
 \label{fig:Pi_y_model}
\end{figure}

As mentioned before, the results for the fit parameters can be used to generate synthetic data for any desired value of $Q^2$ in the large-$Q^2$ tail of $G_E(Q^2)$. This allows us to compute the position space form factor $\Pibar(y,\qmax)$ for any value of $\qmax$ substituting missing lattice data in the Fourier transform by synthetic data from a given fit model. Typical results from this procedure are shown in Fig.~\ref{fig:Pi_y_model}. Regardless of the model, the signal becomes increasingly peaked at $y=0$ for larger values of $\qmax$ while the large-$y$ tail becomes flatted out, which is the expected behavior.  Moreover, including data generated from the $1/\log$ fit model enhances this effect further at large values of $\qmax$ compared to the dipole or Galster-like fit. \par

\begin{figure}[htp!]
 \centering
  \subfigure{\includegraphics[totalheight=0.2\textheight]{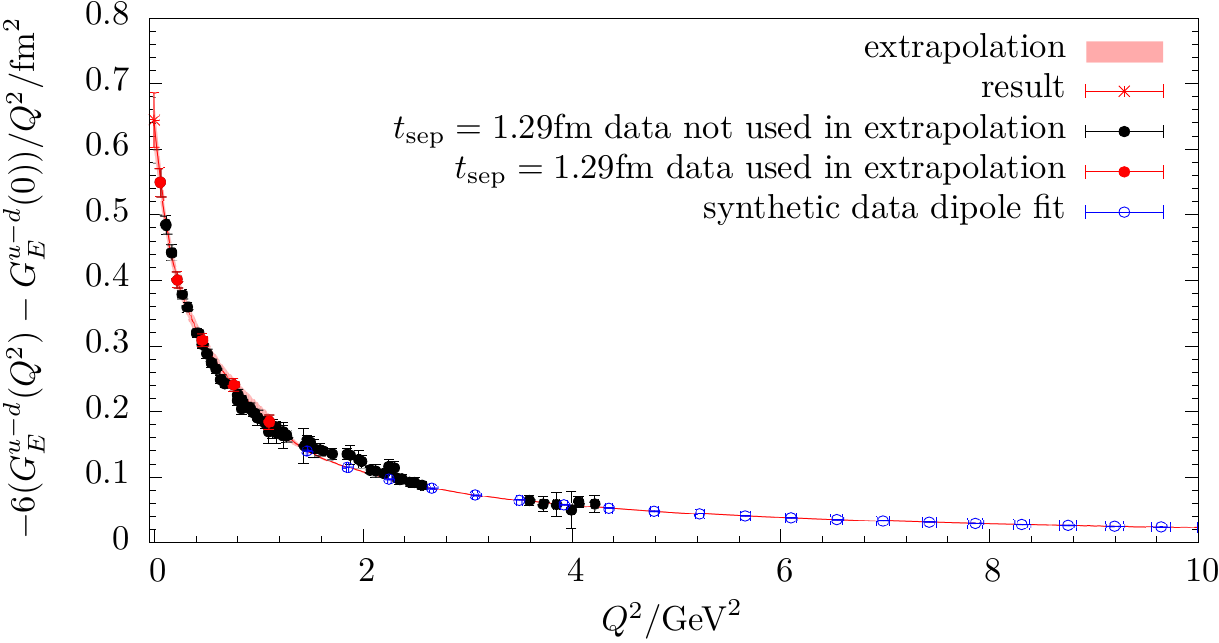}}
  \subfigure{\includegraphics[totalheight=0.2\textheight]{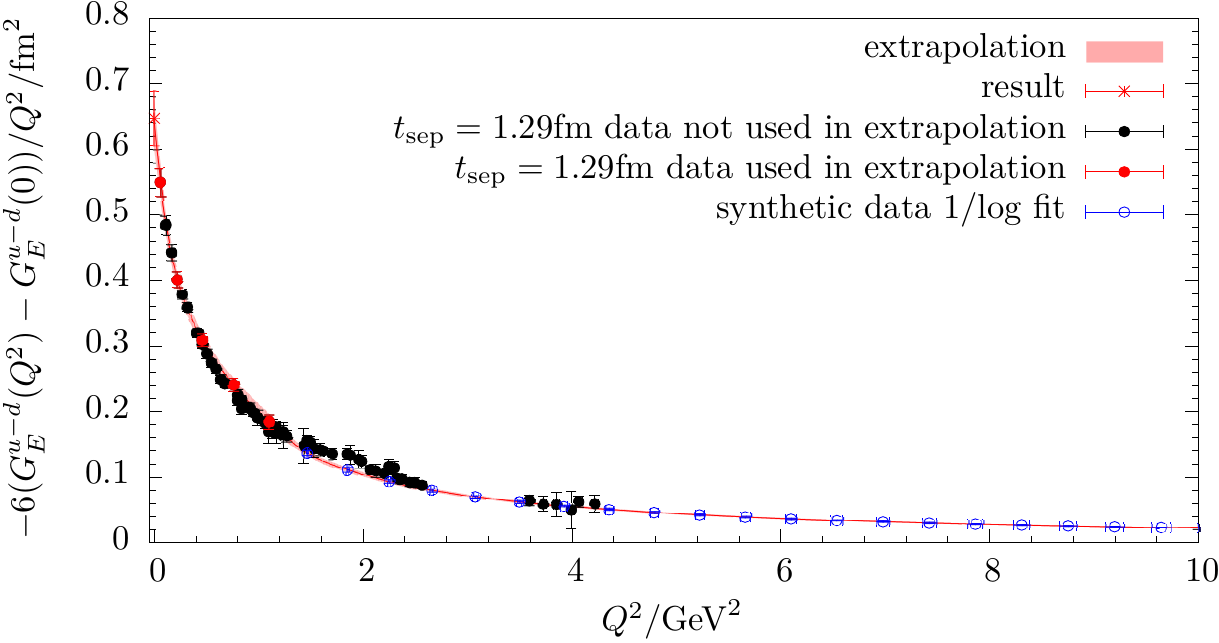}} \\
  \subfigure{\includegraphics[totalheight=0.2\textheight]{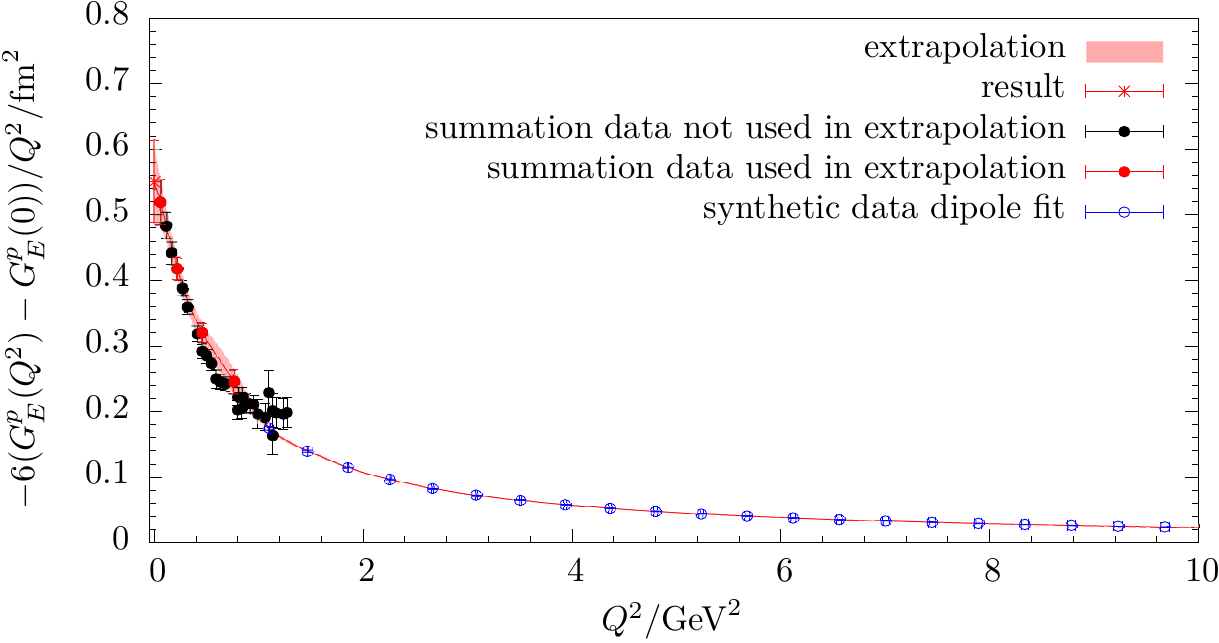}}
  \subfigure{\includegraphics[totalheight=0.2\textheight]{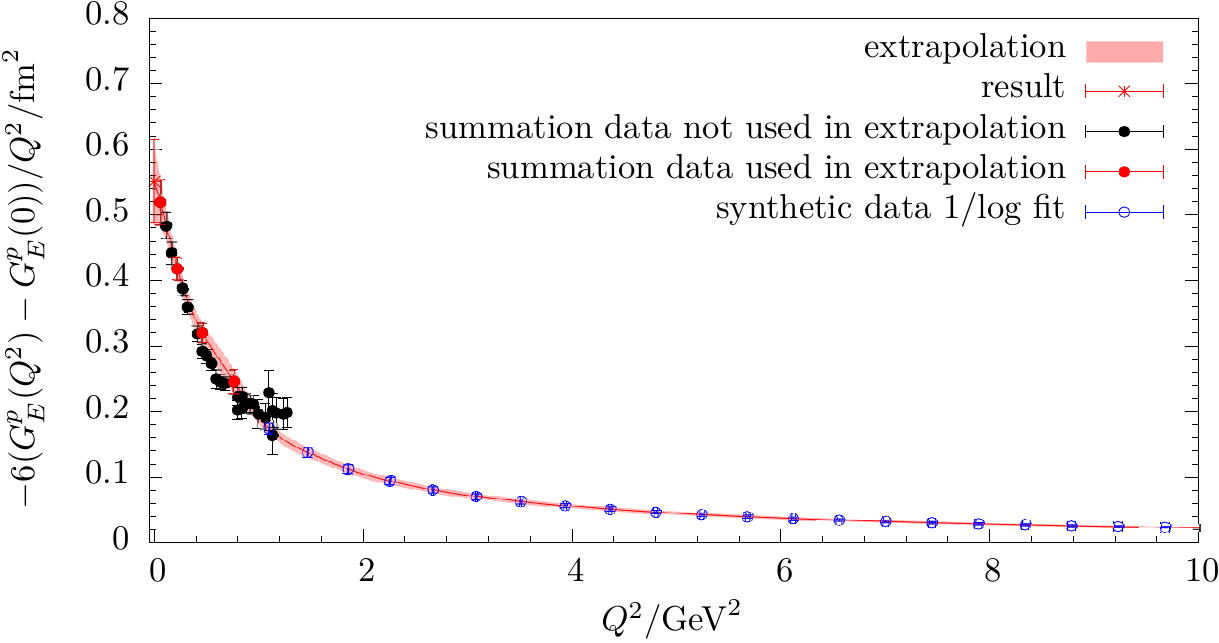}} \\
  \subfigure{\includegraphics[totalheight=0.197\textheight]{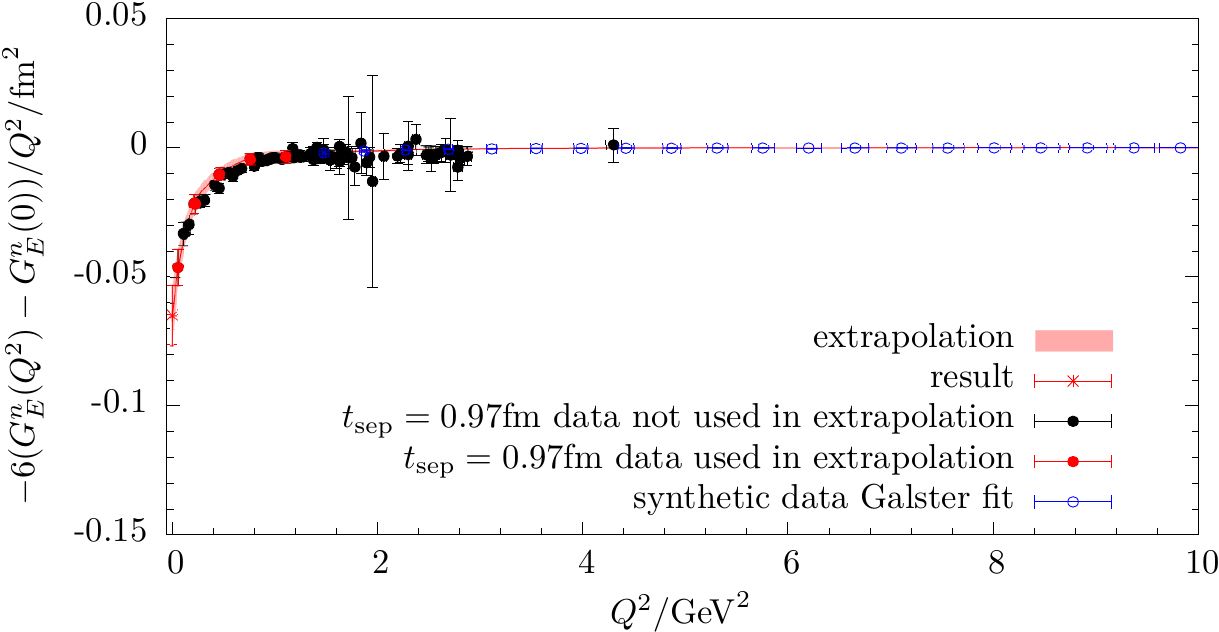}}
  \subfigure{\includegraphics[totalheight=0.197\textheight]{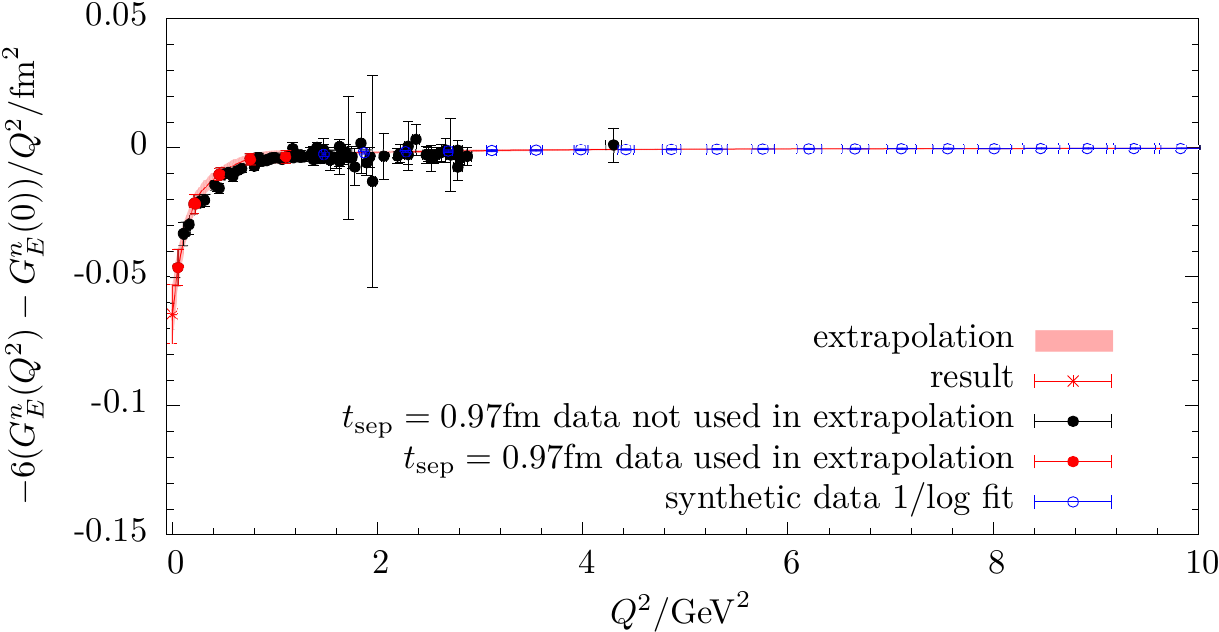}}
 \caption{Lattice (filled symbols) and synthetic data (open symbols) at large $Q^2$ for $-6 (G_E(Q^2)^{u-d}-G_E^{u-d}(0))/Q^2$ together with the continuous extrapolation band computed from Eq.~(\ref{eq:master_formula}). The first, second, and last rows show results for the isovector data at $\tsep=1.29\fm$, the proton data from the summation method and the neutron data at $\tsep=0.97\fm$, respectively. In the left column the dipole fit and Galster-like fit models have been used in the generation of synthetic data, while for the right column synthetic data from a $1/\log$-fit have been used. For the isovector combination and the neutron, synthetic data have been used for $q\geq6\cdot(2\pi/L)$ in the construction of the extrapolation, while for the more noisy proton data from the summation method the switch to model data occurs at $q\geq5\cdot(2\pi/L)$.}
 \label{fig:rsqr_from_modeled_tail}
\end{figure}

Some results for the final extrapolations together with lattice and model data as a function of $Q^2$ are shown in Fig.~\ref{fig:rsqr_from_modeled_tail} for all three isospin combinations and different fit models. We find that there is hardly any difference visible between the extrapolations obtained from using synthetic data from the dipole or Galster-like fit model in the left panels and the $1/\log$ fit model in the right panels of Fig.~\ref{fig:rsqr_from_modeled_tail}. In the presented examples we have employed model data for either $q\geq6\cdot(2\pi/L)$ or $q\geq5\cdot(2\pi/L)$ depending on whether results have been obtained at finite values of $\tsep$ or from the summation method, respectively. At any rate, this confirms the expectation from the kernel weight function in Fig.~\ref{fig:khat2_eq_0_kernel_weight} that the final result is indeed saturated by the first few values of on-axis momentum $q$. Therefore, residual systematic effects related to the truncation of the Fourier transform must be small. \par 

\subsection{Further systematics and final values}
\begin{figure}[ht!]
 \centering
  \subfigure{\includegraphics[totalheight=0.2\textheight]{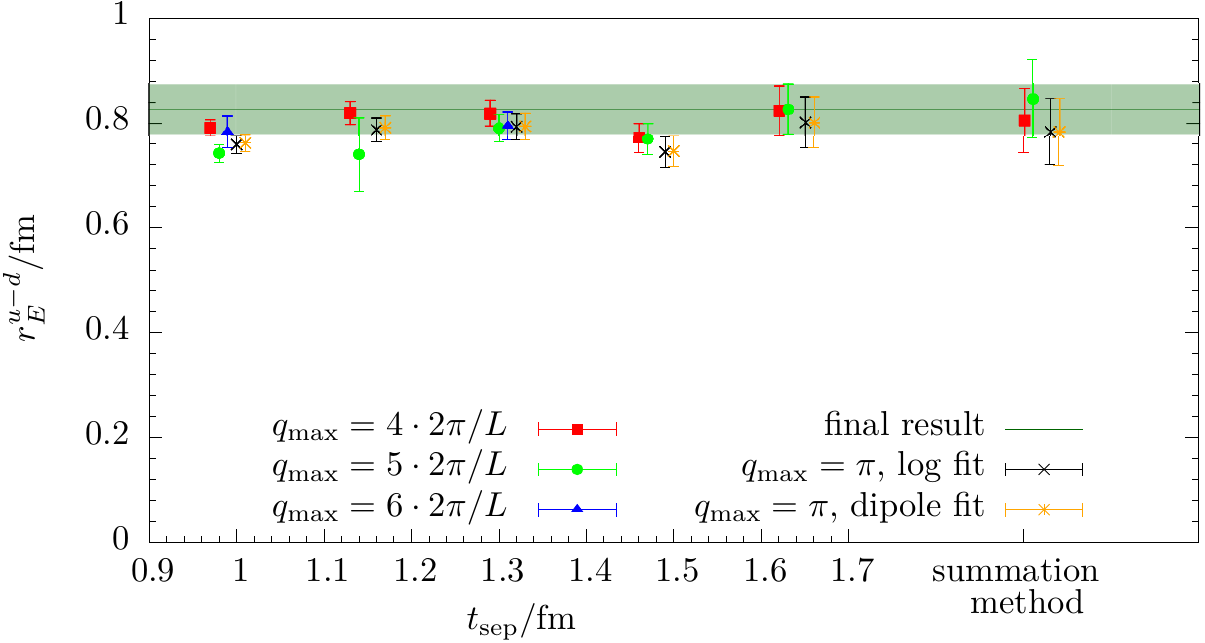}}
  \subfigure{\includegraphics[totalheight=0.2\textheight]{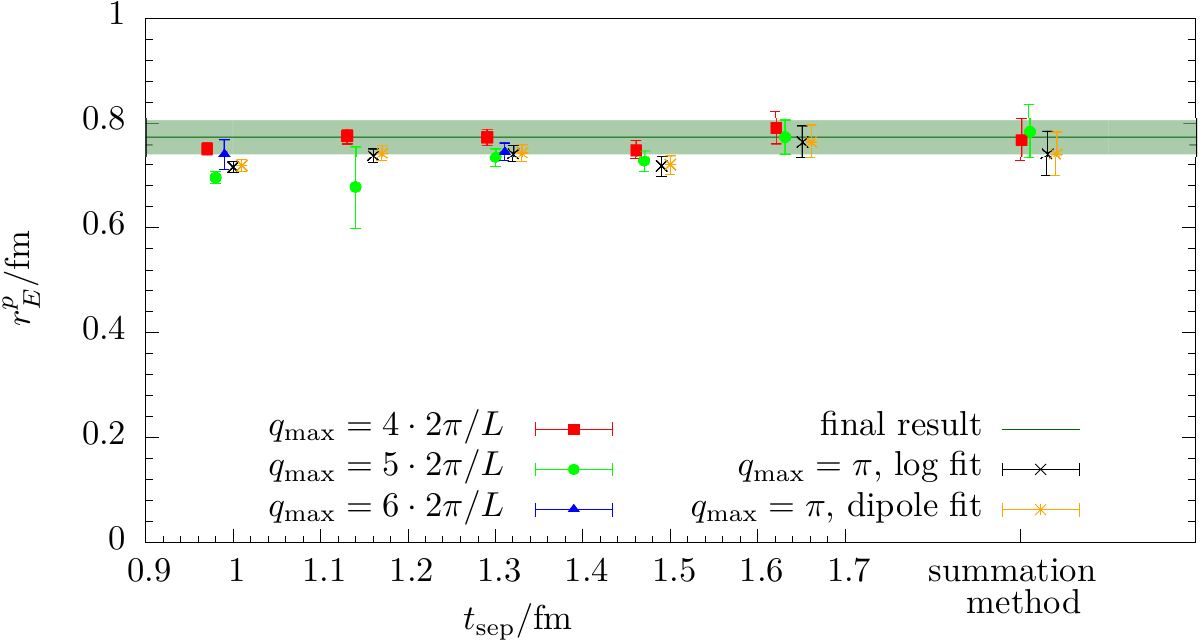}} \\
  \subfigure{\includegraphics[totalheight=0.2\textheight]{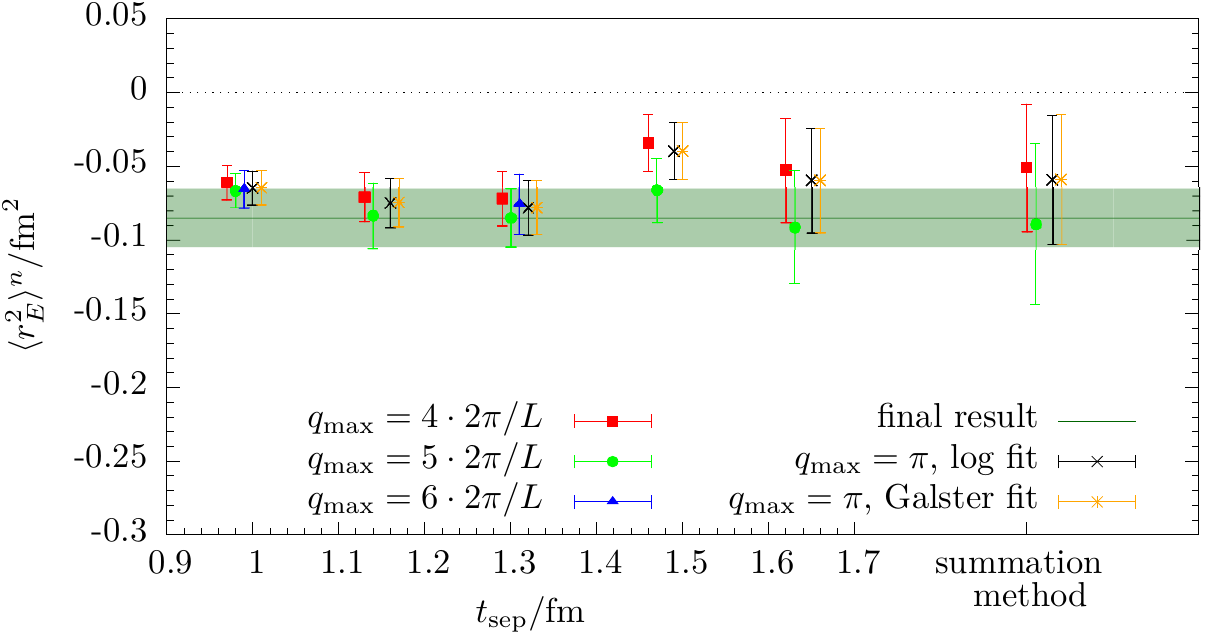}}
  \caption{Overview of results for $r_E^{u-d}$, $r_E^p$ and $\langle r_E^2\rangle^n$ in physical units. Results are shown as a function of $\tsep/a$ and for the summation method using different values of $\qmax$ as well as for two possible choices of modeling the large-$Q^2$ tail of the lattice data for each isospin combination. Results shown for $\qmax/(2\pi/L) = 4,5,6$ use only lattice data, while results for $\qmax=\pi$ use model data for $q\geq 5 \cdot (2\pi/L)$.}
 \label{fig:overview_results}
\end{figure}

The smallness of systematic effects due to truncation of the Fourier transform is further corroborated by a direct comparison of the results for $r_E^{u-d,p}$ and $\langle r_E^2\rangle^n$ at different cutoffs $\qmax$ as well as using synthetic data from the fit models. An overview of results for all available values of $\tsep$ together with the summation method is shown in Fig.~\ref{fig:overview_results}. The corresponding numerical values are listed in Table~\ref{tab:radii} together with the mean-squared radii for the proton and isovector combination. In general, the results using data from the two fit models are in excellent agreement at any given value of $\tsep$. A significant difference is only observed between the two smallest values of the cutoff $\qmax=4\cdot(2\pi/L)$ and $\qmax=5\cdot(2\pi/L)$ for the statistically most precise data at the smallest three source-sink time separations. However, the results obtained with a cutoff of $\qmax=5\cdot(2\pi/L)$ in the Fourier transform are compatible with the results obtained from modeling the large-$Q^2$ tail of the form factor. \par

\begin{table}[!t]
 \centering
  \begin{tabular}{cccccc}
   \hline\hline
   $\tsep/\mathrm{fm}$ & $\langle r^2_E \rangle^{u-d}$ & $\langle r^2_E \rangle^{p}$ & $\langle r^2_E \rangle^{n}$ & $r_E^{u-d}$ &  $r_E^{p}$ \\
   \hline\hline
   0.97      & 0.551(25)  & 0.484(16) & --0.067(12) & 0.742(17) & 0.696(11) \\
   1.13      & 0.55(11)   & 0.46(11)  & --0.083(22) & 0.740(71) & 0.678(78) \\
   1.29      & 0.626(42)  & 0.541(26) & --0.085(20) & 0.791(26) & 0.735(18) \\
   1.46      & 0.593(46)  & 0.530(29) & --0.066(22) & 0.770(30) & 0.728(20) \\
   1.62      & 0.683(78)  & 0.599(50) & --0.091(38) & 0.827(47) & 0.774(32) \\
   \hline
   Summation & 0.72(13)   & 0.616(79) & --0.089(55) & 0.846(74) & 0.785(51) \\ 
   \hline\hline
  \end{tabular}
  \caption{Results for $\langle r^2_E \rangle$ in physical units for the isovector combination as well as the proton and the neutron measured only from lattice data for all five available values of $t_\mathrm{sep}$ and the summation method. For the isovector combination and the proton results for the root-mean-square radius are included as well. All results have been obtained for $\qmax=5 \cdot (2\pi/L)$ using only actual lattice data. Errors are statistical only.}
 \label{tab:radii}
\end{table}

Regarding excited state contamination, there is a weak trend visible for the first few values of $\tsep$. We find that for the proton and isovector radius the results at the largest available value of $\tsep=1.62\fm$ are in good agreement with the summation method. Therefore, we quote the value obtained at the largest available value of $t_\mathrm{sep}$ and for $\qmax=5\cdot (2\pi/L)$ as our final results for the isovector and proton radius, i.e.,
\begin{align}
 r_E^{u-d} &= 0.827\staterr{47}(05)_\mathrm{a}\fm \,, \label{eq:r_iso} \\
 r_E^{p}   &= 0.774\staterr{32}(04)_\mathrm{a}\fm \,, \label{eq:r_p}
\end{align}
where the first error is statistical and the second error refers to the scale setting uncertainty. Note that all final results are obtained from using lattice data only and that synthetic data is only used in the cross-checking the impact of the neglected tail contribution. At the current level of precision, the statistical error clearly dominates and systematics due to the scale settings are negligible. The data for the neutron combination are more noisy and the signal for $\langle r_E^2 \rangle^{n}$ is essentially lost after the third source-sink time separation. For that reason, we quote the value obtained at $t_\mathrm{sep} = 1.29\fm$ and $\qmax=5\cdot (2\pi/L)$ as the final result
\begin{equation}
 \langle r_E^2 \rangle^n = -0.085\staterr{20}(01)_\mathrm{a} \fm \,.
\label{eq:rsqr_n}
\end{equation}

\subsection{Comparison with other studies} \label{subsec:comparison}
Although electromagnetic form factors have been studied within lattice QCD since many years, it is only recently that they have been extracted using simulations with physical values of the light quark masses, referred to as physical point ensembles. Also, while there are a number of studies for the isovector combination, namely, the difference between proton and neutron form factors, for the proton and neutron themselves the results are scarce. This is due to the complexity of evaluating accurately contributions from disconnected quark loops. Therefore, we compare the results of our direct method with recent results extracted using simulations with nearly physical pion masses neglecting the quark-disconnected contributions. Such contributions yield a correction at the 15\% level for the neutron electric form factors, while for the proton the correction is only at the one percent level. We summarize below recent simulations used by various collaborations for the extraction of the nucleon electromagnetic form factors.
\renewcommand{\theenumi}{(\roman{enumi})}
\begin{enumerate}

\item ETMC analyzed three physical point ensembles~\cite{Alexandrou:2018sjm,Alexandrou:2017ypw} of twisted mass fermions: a $N_f=2+1+1$ ensemble with $m_\pi L =3.62$ which has also been used in the present study, and two $N_f=2$ ensembles with $m_\pi L=2.98$ and $m_\pi L =3.97$ and the same lattice spacing of $a=0.0938(3)(1)$~fm. In any case, results are automatically $\mathcal{O}(a)$-improved; hence, cutoff effects are of $\mathcal{O}(a^2)$.

\item LHPC analyzed two $N_f=2+1$ ensembles simulated with two step HEX-smeared clover fermions: an ensemble with $m_\pi =149$~MeV, lattice spacing $a=0.116$~fm, and $m_\pi L=4.21$~\cite{Green:2014xba} and one ensemble at $m_\pi=135$~MeV and $m_\pi L=4$ with a finer lattice spacing of $a=0.093$~fm~\cite{Hasan:2017wwt}. In the latter study they employed a momentum derivative method to extract directly the radii. They use improved currents; therefore, their cutoff effects are ${\cal O}(a^2)$.

\item The PACS Collaboration analyzed two ensembles of $N_f=2+1$ stout-smeared clover fermions: one ensemble of $m_\pi=146$~MeV, $a=0.085$~fm, and $m_\pi L=6$~\cite{Ishikawa:2018rew}, and one ensemble with $m_\pi=135$~MeV, $a = 0.08457(67)$~fm and $m_\pi L=7.4$~\cite{Shintani:2018ozy}. No improved currents have been used; thus, cutoff effects are of ${\cal O}(a)$.

\item The PNDME Collaboration~\cite{Jang:2019jkn} analyzed eleven ensembles of $N_f=2+1+1$ highly improved staggered quarks simulated by MILC Collaboration. They used a mixed action setup with clover fermions in the valence sector. The ensembles have lattice spacings $a \simeq 0.06, 0.09, 0.12, 0.15$~fm and the pion masses are $m_\pi \simeq 135,225,315$~MeV. Using a combined fit analysis, they performed a chiral, continuum and infinite volume extrapolation. We limit ourselves here to their results obtained using the $m_\pi=135$~MeV ensemble with $a=0.0570(1)$~fm and $m_\pi L=3.7$. No improved currents have been used; thus, cutoff effects are of ${\cal O}(a)$.

\end{enumerate}

\begin{figure}[ht!]
  \begin{center}
    \includegraphics[scale=0.5]{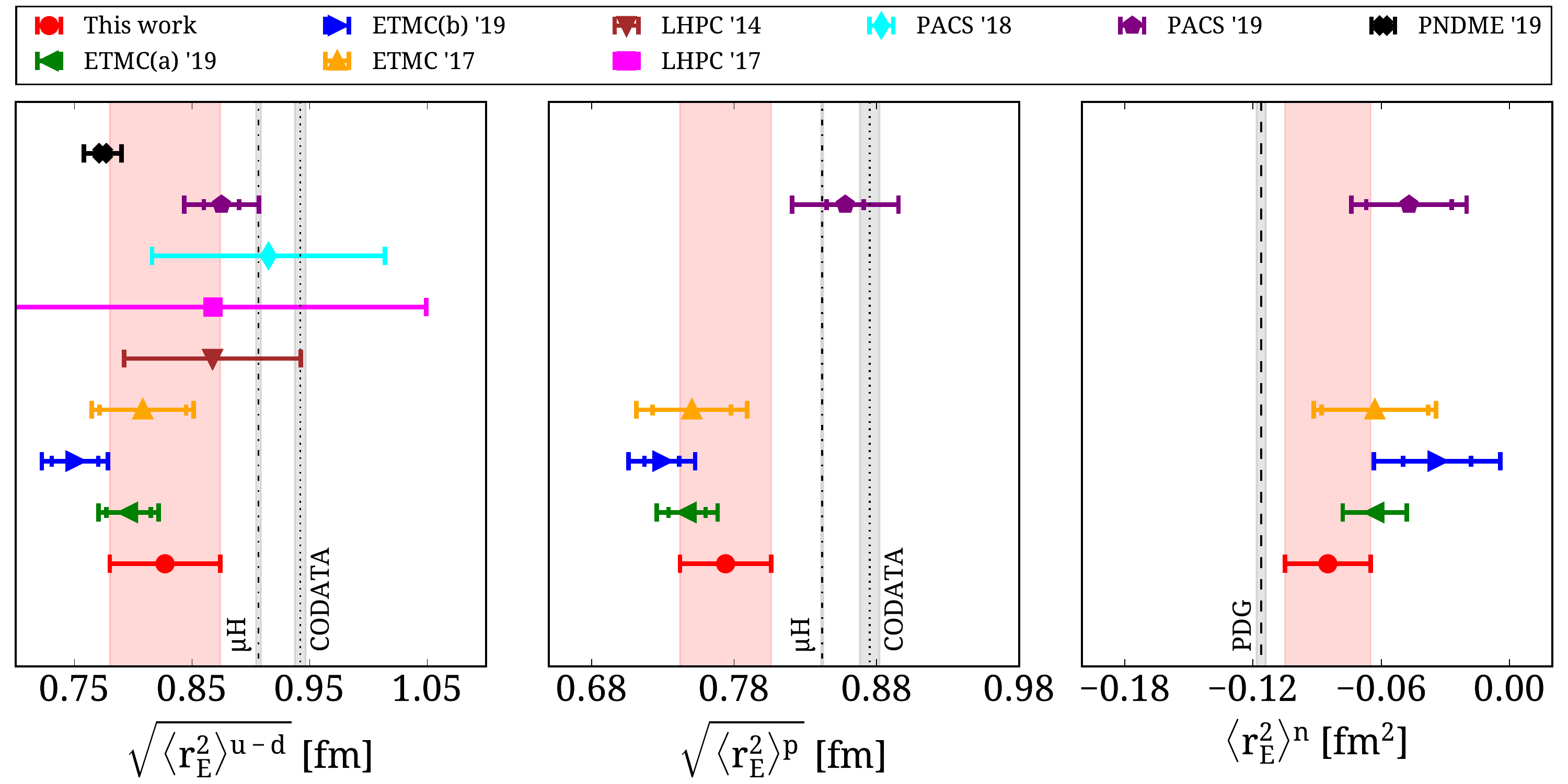}
    \caption{Results from this study (red circles) for $\sqrt{\langle r^2_E \rangle^{u-d}}$ (left panel), $\sqrt{\langle r^2_E \rangle^{p}}$ (middle panel), and $\langle r^2_E \rangle^{n}$ (right panel) compared to results from other lattice QCD analyses. Results are shown for ETMC for the $N_f=2+1+1$ ensemble (green left-pointing triangles), $N_f=2$ and $m_\pi L \simeq 4$ (blue right-pointing triangles)~\cite{Alexandrou:2018sjm}, and $N_f=2$ and $m_\pi L \simeq 3$ (orange upward-pointing triangles)~\cite{Alexandrou:2017ypw}; LHPC using an $N_f=2+1$ ensemble with $m_\pi L=4.21$~\cite{Green:2014xba} (brown downward-pointing triangles) and an $N_f=2+1$ ensemble with $m_\pi L=4$~\cite{Hasan:2017wwt} (square magenta); PACS using an $N_f=2+1$ ensemble with $m_\pi L=6$~\cite{Ishikawa:2018rew} (cyan rhombus) and an ensemble with $m_\pi L=7.4$~\cite{Shintani:2018ozy} (purple pentagon); PNDME~\cite{Jang:2019jkn} using an $N_f=2+1+1$ mixed action and their physical point ensemble with $m_\pi L=3.7$ (black crosses). Note that quark-disconnected diagrams have only been included in the ETMC'17 and ETMC(a),(b)'19 results. The experimental result extracted from the muonic hydrogen~\cite{Pohl:2010zza} is given by the vertical dashed-dotted line and the value referenced by CODATA~\cite{Mohr:2015ccw} by the dotted vertical line. The PDG value~\cite{Tanabashi:2018oca} is given by the dashed vertical line. We note that the PRad~\cite{Xiong:2019umf} experiment finds compatible value for the proton radius with the muonic hydrogen experiment albeit with much larger error; thus, has not been included in the figure.}
    \label{fig:rE_comp}
  \end{center}
\end{figure}
In Fig.~\ref{fig:rE_comp} we compare the results for the radii as extracted from this study with the aforementioned studies. For the isovector combination we find that the value computed within our direct method is compatible with all other studies. We note that one of the three ETMC values is extracted by analyzing the same ensemble but using a dipole fit to extract the radius instead. For the proton, we find very good agreement with most other determinations. Only the study by PACS using a $m_\pi L=7.4$~\cite{Shintani:2018ozy} ensemble obtained a larger value than the rest; however, given the size of errors, the deviation is not statistically significant. Whether this larger value, which is consistent with the experimental determinations, is due to the usage of larger volume needs further investigation. Clearly, however, all lattice QCD results in Fig.~\ref{fig:rE_comp} have errors that are compatible with the difference between the experiment result from the $\mu H$ experiment and that from CODATA. This holds also true for other recent lattice results, e.g., Ref.~\cite{Capitani:2015sba} that has not been included in Fig.~\ref{fig:rE_comp}. Therefore, it is not yet possible to make a distinction between these two values based on lattice results, even if disconnected contributions are included. In fact, disconnected contributions have been included in Ref.~\cite{Alexandrou:2018sjm} and shown to yield a correction of at most $1\%$ for the proton radius which is significantly below the current statistical precision. For the neutron, the relative errors on lattice data are much larger and there is overall agreement among lattice results. The value extracted in this work is in fact very close to the PDG value, despite that quark-disconnected contributions are not yet included.

\section{Summary and Outlook} \label{sec:summary}
In this work we extract the rms charge radii avoiding a fit \emph{Ansatz} to the electric form factors of the nucleon that may introduce a model error. The method has been shown to be insensitive toward the large-$Q^2$ tail of the form factor by testing different \emph{Ans\"atze} to model the large $Q^2$ dependence of the form factors where lattice QCD data are not available. In particular, even using an unphysical \emph{Ansatz} that falls very slowly with $Q^2$ the results are unaffected, demonstrating that lattice QCD results up to about $Q^2\sim 2$~GeV$^2$ determine the radii for the currently available lattice volume and statistical precision. 
Comparing the results for $r^{p-n}_E $, $r_E^p$ and $r_E^n$ extracted by applying this approach to the values 
when the traditional approach of fitting the electric form factors is used for the same ensemble we find consistent values.
Moreover, our values agree with the model-independent determination using the approach of algebraic derivative and quark propagator expansion 
for the direct calculation of form factor derivatives \cite{deDivitiis:2012vs} employed by LHPC \cite{Hasan:2017wwt}. 
In all approaches, the statistical errors on the proton charge radius are still large and lattice results can currently not distinguish between the values extracted from muonic hydrogen and older electron scattering experiments. Besides, most calculations do not yet contain quark-disconnected contribution.

We plan to implement this approach to study the magnetic radii and moments where an increased precision in the lattice QCD data will be required. For the neutron, one has also to include the disconnected contribution which may bring the value closer to the experimental one. Similarly, disconnected diagrams will have to be included if aiming for $\lesssim 2\%$ precision on the proton radius. Finally, nonzero lattice spacing and finite volume artifacts need to be evaluated. This can only take place when ensembles at physical quark mass values are simulated for smaller lattice spacings and larger volumes. Such a program will be possible as these ensembles are simulated and analyzed.

\section*{ACKNOWLEDGMENTS}
We thank all members of ETMC for the most enjoyable collaboration. 
The open source software packages tmLQCD~\cite{Jansen:2009xp} and Lemon~\cite{Deuzeman:2011wz} have been used.
We acknowledge funding from the European Union's Horizon 2020 research and innovation programme under the Marie Sklodowska-Curie Grant No. 642069 and from the COMPLEMENTARY/0916/0015 project funded by the Cyprus Research Promotion Foundation. K.H. is financially supported by the Cyprus Research Promotion foundation under Contract No. POST-DOC/0718/0100. M.P. gratefully acknowledges support by the Sino-German collaborative research center CRC-110. This work was supported by a grant from the Swiss National Supercomputing Centre (CSCS) under Project ID s702. We thank the staff of CSCS for access to the computational resources and for their constant support. The Gauss Centre for Supercomputing e.V. (www.gauss-centre.eu) funded the project pr74yo by providing computing time on the GCS Supercomputer SuperMUC at Leibniz Supercomputing Centre (www.lrz.de).

\end{document}